\documentclass[onecolumn]{emulateapj}
\usepackage{apjfonts}

\newcommand{\bnabla}{\mbox{\boldmath$\nabla$}}
\newcommand{\Om}{\mbox{\boldmath$\Omega$}}
\newcommand{\bmu}{\mbox{\boldmath$\mu$}}

\newcommand{\be}{\begin{equation}}
\newcommand{\ee}{\end{equation}}
\newcommand{\beq}{\begin{eqnarray}}
\newcommand{\eeq}{\end{eqnarray}}
\newcommand{\nn}{\mbox{} \nonumber \\ \mbox{} }
\newcommand{\ba}{\begin{eqnarray}}
\newcommand{\ea}{\end{eqnarray}}

\newcommand{\B}{{\bf B}}

\newcommand\lo{\mathrel{\raise.3ex\hbox{$<$}\mkern-14mu\lower0.6ex\hbox{$\sim$}}}
\newcommand\go{\mathrel{\raise.3ex\hbox{$>$}\mkern-14mu\lower0.6ex\hbox{$\sim$}}}
\def\simlt{\lower.5ex\hbox{$\; \buildrel < \over \sim \;$}}
\def\simgt{\lower.5ex\hbox{$\; \buildrel > \over \sim \;$}}
\def\BQ{B_{\rm Q}}

\def\th{\theta_{\rm em}}
\def\Ethr{E_{\rm thr}}
\def\bB{{\mathbf B}}
\def\RNS{R_{\rm NS}}
\def\BNS{B_{\rm NS}}

\def\K3{\widetilde \kappa_3}
\def\eps3{\widetilde \varepsilon_3}

\def\rhoGJ{\rho_{\rm GJ}}

\def\Tbb{T_{\rm bb}}
\def\kB{k_{\rm B}}

\def\lbar{\rlap{$\lambda$}^{\_\_}_{\;\;\;e}}

\def\Mpm{{\cal M}_\pm}

\def\thkB{\theta_{kB}}
 \def\mukB{\mu}
\def\mumin{\mu_{\rm min}}

\def\Th{\Theta_{\rm bb}}

\def\yparam{\widetilde\omega}

\def\out{\rhd}
\def\in{\lhd}
\def\Gresout{\Gamma^{\rm res}_\out}
\def\Gresin{\Gamma^{\rm res}_\in}
\def\yout{\widetilde\omega_\out}
\def\yin{\widetilde\omega_\in}

\def\gamout{\gamma_\out}

\def\omX{\omega_X}

\def\sigres{\sigma_{\rm res}^0}
\def\alf{\alpha_f}

\shorttitle{ELECTRODYNAMICS OF MAGNETARS: PAIR CREATION AND DYNAMIC
PARTICLE HEATING}
\shortauthors{THOMPSON}
\slugcomment{Submitted to the Astrophysical Journal}
\begin{document}
\title{ELECTRODYNAMICS OF MAGNETARS III:\\
       PAIR CREATION PROCESSES IN AN ULTRASTRONG MAGNETIC FIELD AND\\
       PARTICLE HEATING IN A DYNAMIC MAGNETOSPHERE}

\author{Christopher Thompson}
\affil{CITA, 60 St. George St., Toronto, ON M5S 3H8, Canada}

\begin{abstract}
  We consider the details of the QED processes that create
  electron-positron pairs in magnetic fields approaching and exceeding
  $10^{14}$ G.  The formation of free and bound pairs is addressed, 
  and the importance of positronium dissociation by thermal X-rays is noted. 
  We calculate the collision cross section between an 
  X-ray and a gamma ray, and point out a resonance in the cross section
  when the gamma ray is close to the threshold for pair conversion.  
  We also discuss how the pair creation rate in the open-field
  circuit and the outer magnetosphere can be strongly enhanced by instabilities
  near the light cylinder.  When the current has a strong fluctuating 
  component, a cascade develops.  We examine the details of particle
  heating, and show that a high rate of pair creation can be sustained close
  to the star, but only if the spin period is shorter than several seconds.
  The dissipation rate in this turbulent state can easily accommodate
  the observed radio output of the transient radio-emitting magnetars,
  and even their infrared emission.  Finally, we outline how a very
  high rate of pair creation on the open magnetic field lines can help
  to stabilize a static twist in the closed magnetosphere and to regulate
  the loss of magnetic helicity by reconnection at the light cylinder.
\end{abstract}


\section{Introduction}\label{s:one}

Even when not emitting bright bursts of X-rays, magnetars 
continue to show dramatic variability on much longer timescales.
Much of this variability appears to be a consequence of a dynamic
magnetosphere.  Various manifestations include extremely bright
and persistent hard X-ray emission, which can outstrip the loss
of rotational energy by a thousand times (Kuiper et al. 2004, 2006);
extreme changes in spindown torque that grow and decay over 
a period of months and can, in some cases, be sustained for years
(Kaspi et al. 2001; Woods et al. 2002, 2007; Camilo et al. 2007a,
Camilo et al. 2008);
and, recently, the discovery of high-frequency radio pulsations from two
magnetars (Camilo et al. 2006; 2007b).  A 
general review of magnetars can be found in Woods \& Thompson (2006).

This long-term variability appears to be connected
only indirectly to the sudden, disruptive events known as
Soft Gamma Repeater (SGR) bursts.  Increases in spindown torque in 
the SGRs and some Anomalous X-ray Pulsars (AXPs) typically follow
the X-ray bursts by a period of months.  Very large torque variations
are possible even if the transient X-ray output of the star is
quite modest, as in the radio magnetar XTE J1810$-$197 (Camilo et al. 2007a).
Even more remarkably, given the presence of strong torque variability,
is the strong clustering of the spin periods of most magnetars within
the range of $5-12$ seconds.  The decay of an internal magnetic field,
which ultimately powers the magnetospheric activity,
should not be sensitive to the details of current flow and pair
creation in the magnetosphere.  

Basic energetic considerations suggest that most 
of the dissipation responsible for the very bright, hard X-ray
emission of magnetars is probably concentrated in the 
closed magnetosphere (Thompson \& Beloborodov 2005; Baring \& Harding
2007). The inner magnetosphere of a magnetar 
can sustain much stronger electric currents
than the open-field circuit.  Thompson, Lyutikov, \& Kulkarni (2002) 
pointed out that starquakes create magnetospheric twists that store
an enormous amount of energy in the toroidal magnetic field,
enough to supply the observed non-thermal output for a period of years.
Beloborodov \& Thompson (2007) showed that the voltage 
in the inner magnetosphere is regulated by a continual $e^\pm$ 
discharge.  The electric field that accelerates the particles in
the magnetosphere is sustained by the self-induction of the circuit,
and the loss of energy by radiation induces a gradual decay in the twist.

It has also been suggested that relatively strong 
currents flow within the outer magnetosphere of
a magnetar, being sustained by the outward transfer of magnetic helicity 
from an inner zone where it is injected.  
The redistribution of the twist into the outer magnetosphere causes
the poloidal field lines to expand slightly, thereby increasing
the open-field voltage (Thompson et al. 2002).   This effect depends 
on the extraction of
a tiny fraction the helicity that is stored in the inner magnetosphere,
and therefore does not require a large net input of toroidal field
energy into the magnetosphere.

Considerations of the stability of a twisted magnetosphere,
and observations of strong variations in the brightness of the
radio magnetars on short timescales (Camilo et al. 2007a, 2008),
suggest that the twist in the outer magnetosphere is strongly dynamic.
A first investigation of pair creation in a dynamic magnetosphere
was given by Lyutikov \& Thompson (2005), with application to the
radio eclipses of the double pulsar system J0737$-$3039.   They
noted that strong heating of the particles in the outer magnetosphere
would result from a cascade to high spatial frequencies.  

In this paper we consider the formation of electron-positron discharges 
in the case where the outer magnetosphere is strongly dynamic.
As a first step, we return to first principles and reconsider the 
basic mechanisms of pair creation in some detail.
Previous theoretical work on $e^\pm$ creation and radio death line 
for magnetars was inconclusive. Baring \& Harding (1998, 2001) suggested that 
high-energy photons cannot create pairs because they split in the 
ultrastrong field.   This suppression of pair creation would, however,
take place only if both polarization modes are able to split, a selection rule
which is not supported by microphysical calculations of this process 
(Adler 1971; Usov 2002).   In particular, resonantly 
scattered photons convert very quickly to $e^\pm$ when $B> 4\BQ$.

Magnetars are bright sources of thermal (keV) 
X-rays, and gamma rays are created at a high rate close to the star 
through resonant scattering of the X-rays.  
The scattered gamma rays propagating
through the curved magnetic field can be expected
to convert adiabatically to bound electron-positron pairs 
(Herold, Ruder, \& Wunner 1985; Shabad \& Usov 1986).  We review
this process and identify some peculiar features in super-QED magnetic
fields.  The dissociation of bound pairs
depends on the presence of a source of ionizing photons
(Usov \& Melrose 1996; Potekhin \& Pavlov 1997).  In addition
to photoionization by infrared photons, we
identify the resonant scattering of thermal X-rays by one of
the bound charges as an effective dissociation channel.

We also reconsider the creation of pairs by collisions between gamma
rays and thermal X-rays.  The cross section for this process has
been calculated previously using wave functions for the magnetized
electrons and positrons that are not good eigenstates of the spin
(Kozlenkov \& Mitrofanov 1986).  We recalculate the cross section,
finding a slightly different result when the two colliding photons
are both below the threshold for single-photon pair creation.  We
also identify a near resonance in the collision cross section when
one of the photons is close to the pair creation threshold.

The period clustering of magnetars might,
at first sight, appear to mark out the location
of a radio death line in the $P-\dot P$ plane.  The combined results
of this paper, and a companion paper (Thompson 2008), suggest the
existence of at least two distinct regimes of pair creation:  
a more active regime in which the outer magnetosphere is a 
dynamic reservoir of magnetic helicity and the pair density is
extremely high; and a more quiescent regime in which pair creation occurs
at more modest rate, and the radio emission is concentrated closer to the star.
In this paper, the mechanism of particle heating in the dynamic state is
considered self-consistently, along with the energy and multiplicity 
of the heated pairs. Pair creation
occurs mainly via resonant scattering of thermal X-rays close to the
surface of the star.  A critical spin period arises from the
requirement that the heated charges can emit enough
pair-creating gamma rays to compensate their loss from the circuit.

Many features of the radio magnetars -- their strong flux variability,
hard radio spectra, and broad pulses -- connect them with the 
more dynamic circuit behavior described here.  The voltage that
can be achieved on the more quiescent circuit is examined in 
detail in Thompson (2008), where it is shown that the power dissipated
can barely explain the observed output of the radio magnetars,
but is quite sufficient for more standard pulsar emission.
The spin
clustering of magnetars then signifies a limiting spin period for
the more dynamic and intense regime of pair creation.  Pulsed
radio emission in the more quiescent regime should be
possible at significantly longer spin periods (e.g. Medin \& Lai 2007).

The plan of the paper is as follows.  
We consider the conversion of resonantly scattered photons to pairs
in \S \ref{s:three}, and collisions between gamma rays and thermal
X-rays in \S \ref{s:gxcol}.  The formation of pairs on a magnetic
flux rope with a dynamic twist is treated in   \S \ref{s:six}.
An explanation is given in \S \ref{s:stable} for which current driven 
instabilities in the outer magnetosphere could be regulated by 
a high rate of pair creation on the open magnetic field lines.
Possible sources of short-timescale variations in the open-field
voltage  are examined in \S \ref{s:eight}.  It is shown in \S \ref{s:seven}
that most of the accumulated spindown of a magnetar can occur when
its magnetosphere is in the active state.  The paper closes with
a summary of our results in \S \ref{s:summary}.  The Appendix provides
details of the calculation of the matrix elements for 
$\gamma + \gamma \rightarrow e^ + + e^-$ in an ultrastrong magnetic field.

\section{Creation of Pairs Via Resonant Scattering of Thermal 
             X-rays}\label{s:three}

The voltage on the open magnetic field lines of a neutron
star (and in the closed magnetosphere if it is
non-potential) is regulated by pair creation.  We focus in this
section and \S \ref{s:gxcol} on the creation of pairs by
energetic gamma rays.  The background magnetic field will generally
be assumed to be comparable to, or stronger than, the QED field
\be
\BQ = {m_ec^3\over e\hbar} = 4.4\times 10^{13}\;\;{\rm G}.
\ee

Magnetars are bright thermal X-ray sources, which means that scattering
of the thermal X-rays is the dominant source of pair-creating gamma rays.
This process has been discussed extensively in the context of radio pulsars,
by Kardashev et al. (1984), Daugherty \& Harding (1989), Chang (1995),
Sturner 1995), and Hibschman \& Arons (2001a,b).  Macroscopic electric
fields are screened rapidly enough that the electrons and positrons 
in the inner magnetosphere do
not reach the energies that are required for gamma ray emission by 
curvature radiation.  The electrostatic gap model investigated by
Hibschman \& Arons (2001a,b) and by Thompson (2008) has this property,
as does the more dynamic circuit model described in this paper.

We assume that the target
X-rays are approximately described by a Planck spectrum with temperature 
$\Tbb$, that is, by a spectrum
with an exponential cutoff at $\hbar\omX\gg 3\kB\Tbb$.
This assumption is consistent with measurements of the 
X-ray spectrum of XTE J1810$-$197 (Halpern \& Gotthelf 2005;
Gotthelf \& Halpern 2005).
The spectra of other, brighter, magnetars have a significant tail 
at 1-10~keV, well above the typical $\kB\Tbb=0.4$~keV. However, 
this tail is likely generated at larger radii 
(Thompson et al. 2002; Lyutikov \& Gavriil 2006; Fern\'andez \& 
Thompson 2007) and diluted near the star compared with the thermal 
radiation from the star surface.

Denote the angle of photon propagation with respect to $\bB$ by 
$\thkB$ and let $\mukB=\cos\thkB$. 
For definiteness, let us assume that the magnetic field is predominantly
radial near the star.
The stellar radiation at a radius $r$
occupies the range of angles $\mumin\leq\mukB\leq 1$ where
\be\label{eq:mumin}
  \mumin(r)=\left(1 - {\RNS^2\over r^2}\right)^{1/2}.
\ee 
and $\RNS$ is the stellar radius.
We neglect here the effects of gravitational light bending.
The energy $\hbar\omega$ of a target photon is Doppler-boosted in the 
rest frame of the electron to 
\be\label{eq:loren}
  \hbar\omega^\prime = \gamma(1-\beta\mukB)\,\hbar\omega.
\ee
In this frame, the target photon is strongly aberrated
and moves nearly parallel to $\bB$.   

The scattering cross section is strongly enhanced when the electron
is promoted to its first Landau state.  The resonant
frequency\footnote{The resonant frequency is
given by the non-relativistic cyclotron formula even in 
super-QED magnetic fields, because the absorbing charge 
experiences a recoil along ${\bf B}$.  Note also that
a photon incident along
the magnetic field can be resonantly scattered only at a single frequency,
corresponding to the transition between the lowest and the first Landau 
states.  Resonant interactions involving higher 
Landau states drop out for zero inclination angle.}
in the electron rest frame is (e.g. Herold 1979)
\be\label{eq:resfreq}
   \omega^\prime = {eB\over m_ec}. 
\ee
This translates into a condition on the energy of
target photons in the lab frame, which we write as
\be
\label{eq:res}
\yparam\equiv\frac{\hbar\omega}{\kB\Tbb}=
    \frac{B/\BQ}{\Th\,\gamma(1-\beta\mukB)},
\ee
where 
\be
  \Th\equiv\frac{\kB\Tbb}{m_ec^2}
\ee
is the dimensional temperature of the stellar radiation.
Setting $\beta =1$ is a good approximation if the resonant
scattering occurs close to the star, where $\hbar eB/m_ec \gg \kB\Tbb$.

The target photons move nearly parallel to $\bB$ in the rest frame 
of the scattering particle. 
For the cross section calculation, one can adopt the description 
of resonant scattering as absorption followed by de-excitation. 
Then the cross section is given by (Daugherty \& Ventura 1978),
\be\label{eq:sigres}
  \sigma_{\rm res} = 
  \sigres\,\omega^\prime\delta\left(\omega^\prime-\omega_B\right) = 
  \sigres\,\omega\;\delta\left[\omega-\frac{\omega_B}{\gamma(1-\mukB)}\right],
\ee
where $\omega_B=eB/m_ec$ and
\be
  \sigres = \frac{2\pi^2e}{B}.
\ee
The magnitude of the scattering cross section integrated through the
resonance is 
\be
\langle\sigma_{\rm res}\rangle \;\sim\; \sigres \;\sim\; {3\pi\over 4\alf}\,
\left({B\over\BQ}\right)^{-1}\sigma_T \;\gg\; \sigma_T,
\ee
where $\alf \simeq 1/137$ is the fine structure constant.
Note that $\sigma^{\rm res}$ 
is much larger than the Thomson cross section $\sigma_T$ even in
a $\sim 10^{15}$ G magnetic field.  By comparison, the cross section below the
resonance is strongly suppressed,
\be
\sigma^{\rm non-res} \sim {\sigma_T\over\gamma^2}.
\ee
The two expressions here apply to both X-ray polarization eigenmodes.  These
modes are linearly polarized close to the star, where vacuum polarization
effects dominate the dielectric properties of the medium.  

The rate of resonant scattering is high over a broad range of particle
kinetic energies, due to the broad band nature of the black body radiation
field.  An electron moving radially away from the star
with a Lorentz factor $\gamout\gg 1$ scatters the thermal X-rays at the
rate
\be\label{eq:scatrate}
  \Gresout = \int d\omega \int_{\mumin}^1 2\pi\,d\mukB
  \,\frac{I_\omega}{\hbar\omega}\,(1-\mukB)\,
  \sigma_{\rm res}(\omega,\mukB,\gamout),
\ee
where $\sigma_{\rm res}$ is the scattering cross section, and
$I_\omega$ is the spectral intensity of thermal X-rays, 
\be
\label{eq:bb}
   I_\omega = \frac{(\kB\Tbb)^3}{8\pi^3 c^2\hbar^2}\,f(\yparam),
\ee
\be
\label{eq:f}
    f(\yparam)
      =\frac{\yparam^3}{e^{\yparam}-1}, 
   \qquad \yparam\equiv\frac{\hbar\omega}{\kB\Tbb}.
\ee
We count only one polarization mode in $I_\omega$
as is appropriate to thermal emission from 
a strongly magnetized neutron star (e.g. Silantev \& Iakovlev 1980).

In the case of an outgoing electron near the neutron star
surface, the double integral in equation~(\ref{eq:scatrate}) then yields 
(Sturner 1995)
\be\label{eq:Gresout}
\Gresout = {\alf\Th^3\over 2(B/\BQ)}(1-\mumin)^2\yout^2\left|\ln(1-e^{-\yout})\right|
{c\over\lbar},
\ee
where 
\be\label{eq:resout}
\yout(\gamout) \equiv {\hbar\omega\over \kB\Tbb} = 
{B/\BQ\over \gamout\Th(1-\mumin)},
\ee
and $\gamout$ is the Lorentz factor of the scattering charge,
$\alf=e^2/\hbar c$ is the fine structure constant, and $\lbar=\hbar/m_ec$.
The electron resonantly scatters soft X-rays from the Rayleigh-Jeans tail of
the spectrum when its Lorentz factor is higher than
\be\label{eq:gamoutt}
\gamout^\Theta \equiv  {B/\BQ\over(1-\mumin)\Th} = 
2.3\times 10^4\,{B_{15}\over 1-\mumin}
\left({\kB\Tbb\over 0.5~{\rm keV}}\right)^{-1}.
\ee
The corresponding expression for an ingoing charge is
\be
\label{eq:Gresin3}
\Gresin={\alf\Th^3\over B/\BQ}\,\left(1-\mumin\right)\,f(\yin)\,
   \frac{c}{\lbar},
\ee
where
\be
\label{eq:yine}
   \yin = {\hbar\omega_X\over \kB\Tbb} = 
{B/\BQ\over 2\Th\gamma_{\rm res}} = 1.2\times 10^4\,{B_{15}\over
\gamma_{\rm res}}\,\left({\kB\Tbb\over 0.5~{\rm keV}}\right)^{-1}
\ee
and we have made the approximation $r\gg \RNS$.

\subsection{Direct Conversion of Gamma-Rays to Pairs}

Once created in an ultra-strong magnetic field, a gamma ray will 
rapidly convert to $e^\pm$ pairs.  The details of this
process depend sensitively on the strength of the magnetic field.
The interval between gamma ray emission and pair creation is shortened
substantially if the gamma ray is created above the threshold energy for
pair creation.  Photons this energetic are created by resonant
scattering if the magnetic field is stronger than a threshold value, and
are always created at some finite rate by non-resonant scattering.
We call this process ``direct'' pair creation to distinguish
it from the adiabatic conversion process.

Resonant scattering may be viewed as an excitation of particle
to the first Landau level followed immediately by de-excitation. 
The energy of the scattered photon depends on its pitch angle with respect 
to the magnetic field, $\th$.   It is convenient to view this process in the
frame where the intermediate (excited) electron does not move along $\bB$.
In this frame, the energy of scattered photon is given by 
(Beloborodov \& Thompson 2007),
\be
 \label{eq:th_em}
  E_\gamma(\th)=\frac{E_B}{\sin^2\th}
     \left[1-\left(\cos^2\th+\frac{m_e^2c^4}{E_B^2}
        \,\sin^2\th\right)^{1/2} \right],
\ee  
where $E_B$ is the energy of the first Landau level, 
\be
  E_B=\left(1+2{B\over\BQ}\right)^{1/2}m_ec^2.
\ee
The scattered photon will directly convert to an $e^\pm$ 
pair if its energy exceeds the threshold 
\be 
\label{eq:thresh}
  E_\gamma>\Ethr=\frac{2m_ec^2}{\sin\th}. 
\ee
Photons scattered with 
$\th=\pi/2$ have the minimum $\Ethr=2m_ec^2$ and the maximum 
$E_\gamma=E_B-m_ec^2$ (because the de-exciting electron experiences no 
recoil). The condition $E_\gamma>\Ethr$ is satisfied in a range of 
angles $\th$ if $E_B-m_ec^2>2m_ec^2$, which is equivalent to $B>4\BQ$.
If the surface magnetic field is stronger than this, a large fraction of 
resonantly scattered photons will directly convert to $e^\pm$ near
the surface.  
There is, of course, no constraint on the 
magnetic field for direct pair creation via non-resonant scattering.

\subsection{Adiabatic 
Conversion of a Single Gamma Ray to an Electron-Positron Pair}\label{s:pos}

A gamma-ray that is formed with an energy
$\hbar\omega \gg 2m_ec^2$ and pitch angle $\theta_{kB} \ll 2m_ec^2/\hbar\omega$
is initially below threshold for pair creation, but can propagate
through the curved magnetic field
to a position where the condition (\ref{eq:thresh}) is satisfied
(Sturrock 1971).  This process is familiar from
models of ordinary pulsars with fields $\BNS=10^{12}-10^{13}$~G.
Pair creation by a single photon can occur just above the threshold energy
(\ref{eq:thresh}) only if the magnetic field exceeds a minimum strength,
\be
B \simeq {4\,\BQ\over
3\ln[0.2\alpha_{\rm em}\,\theta_{kB}\,(r m_e c /\hbar)]}
= 0.05\,\BQ \simeq 2\times 10^{12}\;\;{\rm G}.
\ee
(e.g. Erber 1966; Berestetskii et al. 1982).  The corresponding maximum radius for
the creation of a pair in the lowest Landau state is
\be\label{eq:rggmax}
{R_{\gamma\rightarrow e^\pm}\over \RNS} \simeq 8\,B_{\rm NS,15}^{1/3}.
\ee
Pair creation by energetic photons, $E_\gamma \gg 2m_ec^2$,
can occur above threshold in weaker magnetic fields, which leads
to the emission of multiple synchrotron photons.  

In somewhat stronger magnetic fields, the gamma-ray is energetically
favored to convert to bound positronium.  We highlight here some
ambiguities in the existing treatments of this process
(Wunner et al 1985; Shabad \& Usov 1986) when $B \ga \BQ$.
The associated refractive effects have been studied by Shabad \& Usov (1984),
who argue that the group velocity of the gamma ray will bend and
asymptote to the local magnetic field.
The perpendicular energy $\hbar\omega \sin\theta_{kB}$ would, in the process,
become very close to $2m_ec^2$.  Qualitatively,
a gamma ray that enters the dispersive regime at $\hbar k_\perp \sim
2m_ec$ can be viewed as a mixture of a pure electromagnetic mode
(a photon) and a second mode (an electron-positron pair) that
does not propagate across the magnetic field.

The dispersion relation of the gamma ray receives a 
divergent correction from vacuum polarization near the threshold
for pair creation.  Neglecting the interaction between the electron and positron, 
one finds (Shabad \& Usov 1984)
\be\label{eq:disp1}
\omega^2 = c^2k^2 - 2\alf\left({B\over \BQ}\right) 
    \exp\left[{-(k_\perp c/\omega_m)^2\over 2(B/\BQ)}\right]\,
   {\omega_m^3 \over (4\omega_m^2-\omega^2 + c^2k_\parallel^2)^{1/2}},
\ee
where $\omega_m \equiv m_ec^2/\hbar = 2\pi c/\lbar$.
This dispersion relation applies to the two polarization modes which
are luminal at low and high frequencies, and correspond to a propagating
O-mode photon.  The lower branch bends over
from $\omega \simeq ck$ at small momentum, to $\omega \simeq m_ec^2/\hbar$
when $k_\perp \ga m_ec/\hbar$.  

The electrostatic force between the electron and positron will reduce the
energy threshold for pair creation,
\be
E_{\rm thr} \rightarrow {2m_ec^2-E_{\rm bind}(k_\perp,B)\over\sin\theta_{kB}}.
\ee
The binding energy $E_{\rm bind}$ of a positronium atom 
deviates significantly from the zero-field
value ${1\over 2}\times 13.6$ eV when $B > B_{\rm atomic} = \alpha_{\rm em}^2
\BQ = 2.3\times 10^9$ G,
\be\label{eq:ebind}
E_{\rm bind} \simeq {7\over 96}\alpha_{\rm em}^2\,m_ec^2\,\ln^2\left[{B/B_{\rm atomic}
  \over 1 + (B/\BQ)^{-1}(\hbar k_\perp/m_ec)^2}\right]
\ee
(Wunner \& Herold 1979; Shabad \& Usov 1986; Lieb et al. 1992).    The effect of
this interaction on 
the gamma ray dispersion curve have been analyzed by Wunner et al. (1985) and 
Shabad \& Usov (1985) by calculating the vacuum polarization using the positive 
energy wave functions of the bound electron and positron in the magnetic field. 

There are, however, some ambiguities in this procedure when $B>\BQ$.  First,
the dispersion curve (\ref{eq:disp1}) deviates strongly from that of a pure
photon at the wavenumber $k_\perp = 2m_ec/\hbar$.  Writing
$\hbar\omega = 2m_ec^2 - \Delta E$ on the lower branch, one finds
\be\label{eq:dom}
{\Delta E \over 2m_ec^2} \simeq
{1\over 8} \left(2\alf{B\over \BQ}\right)^{2/3}\,\exp\left[-{4\over 3(B/\BQ)}\right].
\ee
One notices that $\Delta E$
has a strong exponential dependence on $\B$, whereas $E_{\rm bind}$
varies only logarithmically.   In an ordinary pulsar-strength magnetic
field, e.g. $B = 4\times 10^{12}$ G, equation (\ref{eq:dom}) gives $\Delta E = 6\times 10^{-4}$
eV, much smaller than the positronium binding energy of $\sim 100$ eV.  The
electron-positron interaction is clearly important in this case, 
and the adiabatic conversion to a bound positronium atom is very plausible.
In a super-QED magnetic field, the energy shift $\Delta E$ is much larger than
the positronium binding energy as calculated from eq. (\ref{eq:ebind}):
one finds $\Delta E = 15$ keV when $B = 4\BQ$.
The energy levels of bound positronium, as calculated in
the Bethe-Salpeter approximation, do not show the same distortion
when $\hbar k_\perp \simeq 2m_ec$ (Shabad \& Usov 1986).   

Second,
the propagation of a positronium atom in a curving magnetic field is
qualitatively different from that of a photon.  
This process is allowed kinematically because the 
generalized momentum of each charged particle contains a term
$e{\bf A}/c$, where ${\bf A} = {1\over 2}\B\times{\bf r}$ is the 
background vector potential and ${\bf r}$ is the displacement from
the center of mass.   The conservation of momentum perpendicular to
$\B$ gives
\be\label{eq:kperp}
\hbar {\bf k}_\perp = {1\over 2}\sum_{\alpha=1}^2 e_\alpha(\B\times {\bf r}_\alpha),
\ee
where $\alpha$ labels the electron and positron ($e_1 = -e$, $e_2 = +e$).  

As the atom moves away from
the star, the electron and positron are individually guided by the magnetic field
and the  perpendicular momentum of the atom {\it decreases}:  equation (\ref{eq:kperp})
implies $k_\perp \sim eB \Delta r_\perp \propto B^{1/2}$.  By contrast, the
perpendicular wavevector of a photon generally increases as the photon propagates
through an inhomogeneous magnetic field:  the geometrical optics equations
imply that ${\bf k}$ is essentially constant (even as the group velocity
of the photon becomes aligned with the magnetic field).

And, third, a
direct calculation of the coulomb correction to the vacuum polarization gives
the same dependence on $B$ (Padden 1994) only if $B<\BQ$.  

In the absence of sufficiently high flux of target X-rays, the 
conversion of a gamma ray to bound positronium remains plausible,
because $\Delta E$ decreases rapidly as $k_\perp$ rises above 
$2m_ec/\hbar$.   Once formed, a positronium atom can be dissociated 
by the resonant scattering of an ambient X-ray at a Landau resonance of
the electron or positron (\S \ref{s:resdis}); 
or by the photoelectric absorption of an ambient IR photon
(\S \ref{s:photo}).  A positronium atom in its ground state is not able to decay
into two photons for a simple kinematic reason (Shabad \& Usov 1982).  
The photon dispersion curve
$\omega(k_\perp)$ flattens near $k_\perp
 = 2m_ec/\hbar$, which means that the conservation of momentum perpendicular to 
${\bf B}$ is not consistent with\footnote{In other words, 
photon splitting is not facilitated 
by strong dispersion near the threshold for pair creation.}
 conservation of energy: 
two daughter photons with a total energy $2m_ec^2-E_{\rm bind}$
must have a smaller momentum than was contained in the atom.   

\subsection{Disintegration of a Positronium Atom through \\  Absorption
of an X-ray at the Electron/Positron Landau Resonance}\label{s:resdis}

Given the ambiguities just mentioned
in the dispersive properties of the hybrid photon-pair,
we approach the problem of pair formation in two ways.  The hybrid
particle is treated either as a pure photon, or a pure positronium atom,
and the relevant pair creation channels are analyzed.
A gamma ray can collide with an ambient X-ray while it is still in the
mixed state.  We find that
the cross section for such a two photon collision is resonantly enhanced when
the gamma ray by itself is close to the threshold for pair conversion
(\S \ref{s:gxcol}).

Although a positronium atom is electrically neutral, the electron and positron
can scatter high-frequency photons whose wavelength is smaller than
the electron-positron separation.  The minimum separation is set by the
transverse momentum of the atom:  $\Delta r_\perp \simeq \hbar k_\perp c/eB$.
A single electron at rest will resonantly absorb a photon of energy $(B/\BQ)m_ec^2$ 
with a cross section (\ref{eq:sigres}).  The corresponding photon wavelength $\lambda$ is
\be
{\lambda\over \Delta r_\perp} = {1\over 2}\,\left({\hbar k_\perp\over 2m_ec}\right)^{-1}.
\ee
We can therefore expect an X-ray photon
to be absorbed by either the electron or 
positron with a cross section close to (\ref{eq:sigres}).  The photon imparts
a recoil momentum to the absorbing particle, which is $\Delta p \sim (B/\BQ)m_ec$
when $B < \BQ$.  The atom is dissociated if $(\Delta p)^2/2m_e \ga E_{\rm bind}$,
which requires $B> 1\times 10^{12}$ G.  Because bound pair creation occurs only
for magnetic fields stronger than $B > 2\times 10^{12}$ G $\sim 0.05\,\BQ$,
dissociation can be assumed to follow immediately following the
resonant absorption of the X-ray.

A positronium atom therefore has a very short mean free path for dissociation
close to an X-ray bright magnetar.  If the original gamma ray is created
by resonant scattering of an X-ray by an electron of Lorentz factor $\gamma$,
then the gamma ray has an energy $\simeq f_{\rm recoil}\gamma m_ec^2$,
where the recoil factor is  $f_{\rm recoil} = {2\over 3}$ when $B = 4\BQ$
and $f_{\rm recoil} \simeq B/\BQ$ for $B \ll \BQ$.
 The resultant positronium atom has a Lorentz factor $\gamma_\pm \simeq 
(f_{\rm recoil}/2)\gamma$.
The original seed photon for scattering, and the photon that dissociates the
positronium atom, therefore tend to be drawn from different
parts of the thermal X-ray peak.  When $B \ga \BQ$, the photons which
act as seeds for pair creation come from above the black body
peak, the dissociating photon comes from close to the peak, and the
positronium is rapidly dissociated.  On the other hand, when $B < \BQ$,
the dissociating photon is drawn from the Rayleigh-Jeans tail and,
if $\kB\Tbb$ is below a critical value, it is possible for a newly 
created positronium atom to avoid dissociation by resonant scattering. 
In this circumstance the circuit voltage could be regulated by
a different pair-creation channel -- such as the direction conversion of
non-resonantly scattered photons to pairs (Thompson 2008) --
even though the creation
rate of gamma rays by resonant scattering were formally higher.

\subsection{Photodisintegration of Parapositronium}
\label{s:photo}

Many magnetars are intense sources of optical-IR
photons, with a characteristic luminosity of $10^{32}$ ergs s$^{-1}$
(Durant \& van Kerkwijk 2006a,b, and references therein). 
These photons are in the
right frequency range to dissociate bound positronium (parapositronium)
when the relativistic motion of the bound pairs along ${\bf B}$
is taken into account.   The binding energy is
$E_0 \simeq 200$ eV in a magnetic field $B \sim 4\,\BQ$ when the
threshold for pair creation is $\hbar k_\perp = 2 m_ec/\hbar$
(eq. [\ref{eq:ebind}]).  Pair-creating gamma rays typically have
energies $\hbar\omega \sim (0.05-1)\gamout^\Theta m_ec^2$ near the
surface of the star, depending
on the details of the circuit model (Thompson 2008).  
The resulting pair has a Lorentz factor 
$\gamma = \hbar\omega/2m_ec^2 \sim (500-10^4)\,B_{15}
(\kB\Tbb/0.5~{\rm keV})^{-1}$ parallel to ${\bf B}$
(see eq. [\ref{eq:gamoutt}]).  Photons of an energy 
\be
\hbar\omega \simeq {E_0\over\gamma(1-\mumin)} \sim (0.02-0.4)
B_{15}^{-1}\left({\kB\Tbb\over 0.5~{\rm keV}}\right)\;\;{\rm eV}
\ee
will be most effective at photodissociating positronium.

The cross section for photodissociation is 
suppressed compared with an
unmagnetized positronium atom (see, e.g., Fig. 1 of Potekhin \&
Pavlov 1997).  The photon is incident along ${\bf B}$
in the rest frame of the atom, and so both (linear) polarization
modes have nearly identical cross sections.  However, the motion of
the charges in the atom is restricted in the plane perpendicular
to ${\bf B}$, and the cross-sectional area of the atom
is reduced by a factor 
\ba
a_B^{-2}\,\left({eB\over \hbar c}\right)^{-1}
 &=& \left({B\over B_{\rm atomic}}\right)^{-1}\nn
&=& 
{\alpha_{\rm em}^2\over 16}\,\left({B\over 4\,\BQ}\right)^{-1}
= 3.3\times 10^{-6}\,\left({B\over 4\,\BQ}\right)^{-1}.
\ea
Here $a_B= 2\hbar^2/m_ee^2$ is the Bohr radius of an
unmagnetized positronium atom, and 
\be
B_{\rm atomic} = {1\over 4}\alpha_{\rm em}^2
\BQ = 6\times 10^8\;\;\;\;{\rm G}
\ee
is the magnetic field that begins
to significantly deform the atom away from its unmagnetized structure.

The cross section for a photon incident parallel to ${\bf B}$
with an energy just above the threshold $E_0$ can be estimated as 
\be
\sigma_{\rm phot}(\omega') \;\sim\; {\pi^2e^2\over (m_e/2)(E_0/\hbar)c}\,
\left({B\over B_{\rm atomic}}\right)^{-1},
\ee
where $m_e/2$ is the reduced mass.  (See Potekhin \& Pavlov 1997 
for more detailed expressions.)
The rate of photodissociation in response to a spectral intensity
$I_\omega$ of optical-IR photons is 
\be
\Gamma_{\rm phot} = \int d\omega\,\int_{\mumin}^1
2\pi\,d\mu\,(1-\mu){I_\omega\over\hbar\omega}
\sigma_{\rm phot}(\omega'),
\ee
as measured in the frame of the star.

We now show that photodissociation is rapid if the optical-IR continuum 
of a magnetar is emitted close to its surface.
In evaluating the
above integral, we also make use of the 
observation that the luminosity density $L_\omega$ is roughly
independent of frequency in at least some AXP sources.  Then
\be
\Gamma_{\rm phot} = (1-\mumin)\,
{\pi e^2\,L_\omega\over 4 r^2 m_ec |E_0|}\,
\left({B\over B_{\rm atomic}}\right)^{-1},
\ee
where $L_\omega/4\pi r^2 = 2\pi I_\omega(1-\mumin)$.
It is convenient to express $\Gamma_{\rm phot}$ in terms of the 
luminosity $\omega L_\omega$ at $\hbar\omega = 0.1$ eV.  
Approximating  $1-\mumin \simeq {1\over 2}(r/\RNS)^{-2}$,
and re-expressing $r$  in terms of the local value of the magnetic field, gives
\be
\Gamma_{\rm phot} = 1\times 10^7\, {(\omega L_\omega)_{32}\over
     B_{\rm NS,15}^{4/3}\,R_{\rm NS,6}^2}\,\left({B\over 4\,\BQ}\right)^{1/3}
\;\;\;\;{\rm s^{-1}}.
\ee

We see that photodissociation is rapid, $\Gamma_{\rm phot} > c/\RNS$,  
 if the surface luminosity at 0.1 eV
is higher than $\sim 10^{29}$ ergs s$^{-1}$.
This critical luminosity is only
$\sim 10^{-2}-10^{-3}$ of the typical K-band luminosities of AXPs; but
it is also $\sim 10^6$ times {\it larger}
than expected from the Rayleigh-Jeans tail of the X-ray black body.
Photodissociation of positronium could therefore be slow if the optical-IR
output were concentrated at a greater distance from the star.
For example, a very dense pair gas forms on the open magnetic field lines
in the circuit model described in \S \ref{s:six}.  The pair density
is high enough to seed
the optical-IR emission of the transient magnetar XTE J1810$-$197
by a form of plasma emission (Thompson 2008).  In that case, the optical-IR
flux would be beamed away from the star, and from the base of the open
magnetic flux tube.

\section{Collisions between Gamma Rays and Thermal X-rays}\label{s:gxcol}

High-energy photons can create pairs by colliding 
with thermal X-rays.  This pair creation channel is important
when the X-ray flux is high (e.g. Zhang \& Qiao 1998).   
Although most gamma rays will first convert to pairs off the magnetic field, 
a significant amount of 
pair creation by photon collisions may take place within a gap.
If the surface temperature of the neutron star is higher than
$\sim k\Tbb=0.2$~keV, then the gap voltage is reduced.

We consider the collision between a gamma ray and a thermal
X-ray that is emitted directly from the magnetar surface,
\be\label{eq:gx}
\gamma + X \rightarrow e^+ + e^-.
\ee
The gamma ray is created with an energy\footnote{In this section,
we employ natural units in which $\hbar = c = 1$.}   $\omega \gg 2m_e$,
but below the threshold for single-photon pair creation,
$\omega \sin\theta_{kB} < 2m_e$.  At the point of pair creation,
it is nearly completely polarized in the ordinary mode, due to  the rapid
splitting of extraordinary mode gamma rays.  The target photon has
an energy $\omega_X \ll m_e$.  The kinematic threshold
for pair creation through the channel (\ref{eq:gx}) is obtained
from the equations of conservation of energy and momentum 
parallel to ${\bf B}$,
\ba
\omega + \omega_X &=& \left(P_+^2 + m_e^2\right)^{1/2} + 
\left(P_-^2 + m_e^2 + 2neB\right)^{1/2};\nn
\mu\,\omega + \mu_X\, \omega_X  &=& P_+ + P_-,
\ea
where $\mu = \cos\theta_{kB}$, $\mu_X = \cos\theta_{kB,X}$ 
are the direction cosines of the gamma ray and X-ray, and
$P_\pm$ are the momenta of the final electron and positron.  One finds
\be
2\omega\omega_X(1-\mu\mu_X) \;>\; (m_e + E_{0n})^2 - \omega^2(1-\mu^2),
\ee
where $E_{0n} = (m_e^2 + 2neB)^{1/2}$ and $\omega(1-\mu^2)^{1/2} = 
\omega \sin\theta_{kB}$ is the perpendicular energy of the gamma ray.

\begin{figure}
\epsscale{0.7}
\plotone{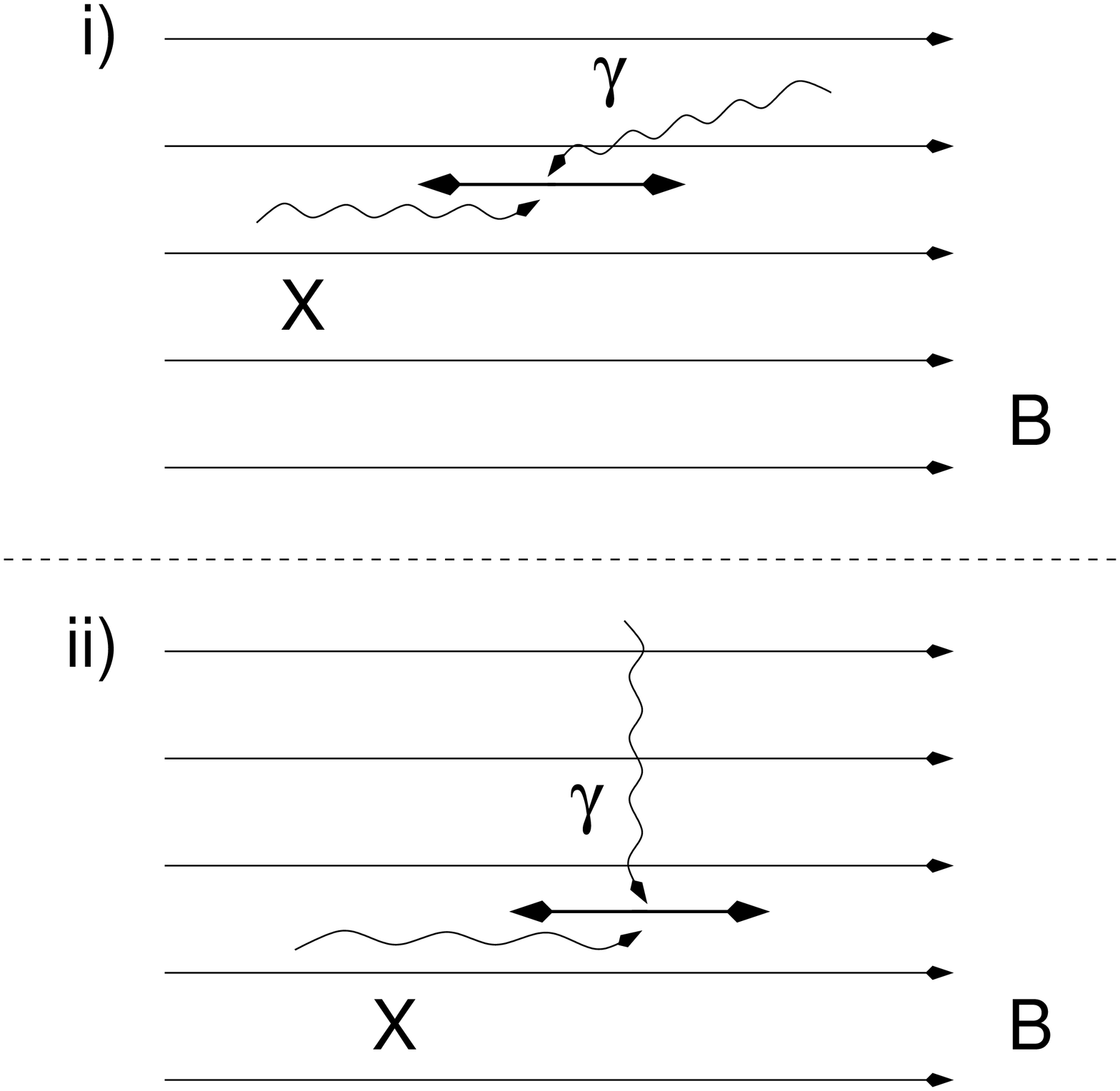}
\caption{Collision of a gamma ray with a target X-ray.
We consider two cases:  i) the collision occurs
close to the point of emission of the gamma ray, so that
the gamma ray is well below the threshold for single
photon pair creation ($\hbar\omega\sin\theta_{kB} \ll 2m_ec^2$);
and ii) the gamma ray is just below threshold
(eq. [\ref{eq:dele}]).  In case i) the two photons
move nearly head on in the center-of-momentum frame
parallel to $\B$.  In case ii), the cross section is
strongly enhanced if the target photon has an energy
$\sim (B/\BQ)m_ec^2$ in the frame where the gamma ray
moves perpendicular to ${\bf B}$.
\vskip .2in\null
}
\label{f:gXcol}
\end{figure}

Two particular cases are of interest here:

\vskip .1in
\noindent
1. ---
The gamma-ray remains well below the threshold for pair creation,
$\omega(1-\mu^2)^{1/2} \ll 2m_e$,
so that vacuum polarization effects introduce very little dispersion
into the relation between $\omega$ and $k$.   The target photon 
is tightly collimated about ${\bf B}$ in the frame where the
net momentum of the two photons along $\B$ vanishes
($\theta_{kB,X}' \sim 2m_ec^2/\hbar\omega$; Fig. \ref{f:gXcol}). 
Pair creation with the both final particles in Landau state $n=0$ has
the same threshold condition
\be\label{eq:threshxg}
\omega \omega_X (1-\mu_X) = 2m_e^2,
\ee
where we have set $\mu \simeq 1$.  This is identical to the threshold
condition in the absence of a magnetic field.  

We have evaluated the cross section for this process to the lowest
order in perturbation theory.  The tree-level Feynman diagrams are
calculated using wavefunctions for the electron and positron that are good
spin eigenstates (Melrose \& Parle 1983a,b). Details are provided in
the Appendix.    We find
\be\label{eq:sigxgth}
\left({V_\pm\over c}\right)\,
\sigma_{X\gamma} = {3\sigma_T\over 16}\,\left({B\over\BQ}\right)
\,\left({E_\pm\over m_e}\right)^{-2}\,
\left|{\varepsilon_2^+\varepsilon_1^- m_e^2\over
(E_\pm + P_\pm)^2 + m_e^2 + 2eB} -
{\varepsilon_1^+\varepsilon_2^- m_e^2\over
(E_\pm - P_\pm)^2 + m_e^2 + 2eB}\right|^2,
\ee
where $E_\pm$ and $P_\pm = P_- = -P_+$ are energy and momenta of
the outgoing electron and positron (both in the lowest Landau state)
in the center-of-momentum frame, and $V_\pm = |P_\pm|/E_\pm$.
The polarization vectors of the two colliding photons 
in this frame are denoted by $\varepsilon^\pm_k = 
\varepsilon^x_k \pm i \varepsilon^y_k$.
Expression (\ref{eq:sigxgth}) simplifies to 
\be\label{eq:sigxgthb}
\left({V_\pm\over c}\right)\,
\sigma_{X\gamma} = {3\sigma_T\over 16}\,
{B/\BQ\over (1+B/\BQ)^2}\,\left(\varepsilon^y_1 \varepsilon^x_2 -
\varepsilon^y_2 \varepsilon^x_1\right)^2\qquad (V_\pm \ll c)
\ee
near threshold.  The coefficient works out to 
$0.03\sigma_T$ in a magnetic field $B = 4\BQ$.  
This result differs somewhat in the dependence on $B$ and the
numerical coefficients from those of Kozlenkov \& Mitrofanov (1986),
who used the wavefunctions for the Dirac particles that are not
good spin eigenstates and did not explicitly evaluate the polarization
dependence.  

\vskip .1in
\noindent 
2. ---
The gamma-ray is just below the threshold for pair creation,
\be\label{eq:dele}
\omega(1-\mu^2)^{1/2} \; =\;  2m_e - \Delta E.
\ee
In the case of adiabatic conversion of a gamma ray to a pair,
$\Delta E$ measures the amount by which the
dispersion curve of the photon bends below the rest energy of
an electron-positron pair.  For example,
$\Delta E \simeq 15$ keV when the perpendicular momentum of the photon is
$k_\perp = 2m_e$ in a magnetic field $B = 4\BQ$ (eq. [\ref{eq:dom}]).

In this situation, there is a very low threshold energy
for the creation of a pair with both particles in Landau level $n=0$.
Transforming to the frame\footnote{Obtained by a boost with
Lorentz factor $\gamma = 1/\sin\theta_{kB} \simeq 2m_e/\Delta E$
along ${\bf B}$.} in which the gamma ray moves perpendicular
to ${\bf B}$, one has
\be
\omega_X' \;>\; \Delta E.
\ee
This energy is even smaller in the stellar frame,
$\omega_X = \omega_X'/\gamma(1-\mu_X)$.  
When one of the final state particles is in a higher Landau
state $n$, one obtains
\be\label{eq:omxp}
\omega_X' = \Delta E + {1\over 2}\left(n{eB\over m_e} + E_{0n} - m_e\right),
\ee
which reduces to $\Delta E + eB/m_e$ in a sub-QED magnetic field.

The cross section is strongly enhanced if the virtual electron that
mediates the collision between the two photons is on mass shell.
This is possible only if the energy of the target X-ray satisfies
a resonance condition.  (The main difference with the resonance
encountered in cyclotron scattering is that the virtual electron
can remain in the lowest Landau state.)  We take the gamma ray
(parallel momentum $\mu'\omega' = 0$) to share a vertex with 
the outgoing positron (see Fig. \ref{f:diagram}).  The parallel momentum
$P_I$ of the virtual electron is related to that of the target
X-ray ($\mu_X'\omega_X' \simeq \omega_X'$) and final-state electron by
\be\label{eq:rescon1}
P_I = -\omega_X' + P_-' = - P_+'.
\ee
The resonance conditions are\footnote{We can equally well exchange
the electron and positron in the final state; both Feynman diagrams
contribute to the resonant cross section.}
\be\label{eq:rescon2}
(\omega_X' - E_-')^2 = (\omega' - E_+')^2 =  P_I^2 + m_e^2.
\ee
We are searching for the lowest-energy resonance, and so take
the virtual electron to be in the lowest Landau level.
The condition of energy conservation 
\be\label{eq:encon}
\omega' + \omega_X' = E_+' + E_-' 
\ee
is satisfied only if one final particle is in the
lowest Landau state, e.g. $n_+ = 0$.
Then the solution to equations (\ref{eq:rescon1}), (\ref{eq:rescon2}) is
\be\label{eq:eeqs}
E_+' = m_e
\qquad
E_-' = \omega_X'\left(1+{m_e^2\over 2n_- eB}\right) + n_- 
   {eB\over 2\omega_X'},
\ee
where we have approximated $\omega' \simeq 2m_e$.
Applying eq. (\ref{eq:encon}) shows that the energy of the
target X-ray must satisfy the same resonance condition as for absorption
by a real charge at its first Landau transition,
\be
\omega_X' = \gamma(1-\mu_X)\omega_X = n_-{eB\over m_ec}.
\qquad (n_- > 0).
\ee
The resonant target photon energy is essentially the same as the
threshold energy (\ref{eq:omxp}) when $B < \BQ$,
but lies substantially above threshold in super-QED magnetic fields.

It should be emphasized that this resonance is only approximate
when $\Delta E > 0$.  Substituting $n_+ = 0$ into eq.
(\ref{eq:eeqs}), one sees that 
\be
E_+' = m_e - {1\over 2}\Delta E.
\ee
A true resonance is not possible, because one particle
in the final state would not be able to satisfy the mass shell
condition. But an approximate resonance is present when $\Delta E \ll 2m_e$.
The cross section derived in the Appendix is
\be\label{eq:sigresxg}
{\sigma\over\sigma_T} \;=\; {3\over 8}e^{-2\BQ/B}
\left({m_e\over \omega'-2E_+'}\right)^2\,
(\varepsilon^z)^2.
\ee
In this expression, 
the frequency $\omega'$ and polarization $\varepsilon^z$
of the incident gamma ray, and the energy $E_+'$ of the outgoing positron,
are evaluated in the frame where the gamma ray perpendicular to $\B$.

The momentum of the positron can be evaluated in 
terms of the displacement of the energies of the gamma ray
and target photon from the resonant values $\omega' = 2m_e$
and $\omega_X' = eB/m_e$:
\be
\left({B\over\BQ}\right)\,P_+' = 
     \left(1+{B\over\BQ}\right)\left(2m_e - \omega'\right)\;+\;
       \omega_X' - {eB\over m_e}.
\ee
The cross section peaks at $P_+' = 0$, where the denominator
of eq. (\ref{eq:sigresxg}) has the minimum value $\Delta E^2$.

The cross section exhibits
additional localized peaks where the energy threshold for excitation
to higher Landau levels is just satisfied (Kozlenkov \& Mitrofanov 1986).
We ignore this effect in this paper, being interested in the case
where the fluxes of gamma rays and target X-rays are both cut off
exponentially at high frequencies.

\section{High Pair Densities in a Turbulent Flux Rope}\label{s:six}

\begin{figure}
\epsscale{1.0}
\plotone{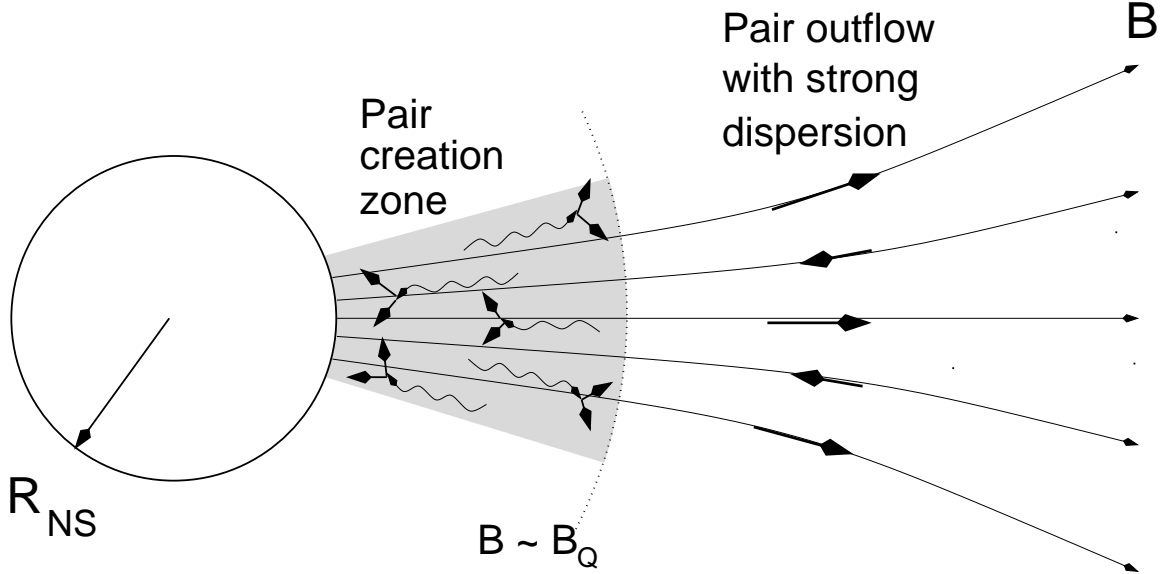}
\caption{
The excitation of a torsional wave on the extended
magnetic field lines (open and closed) heats the 
enclosed electrons and positrons when the wave energy
cascades to a high wavenumber.  The fluctuating component
of the current diverges in the cascade. There is
a critical wavenumber above which the relativistic plasma
particles can no longer conduct charge at the rate demanded
by the waves, and the wave energy is transferred to the 
particles.
When the particle density is much higher than $n_{\rm GJ}$
and the plasma is nearly charge-balanced, then waves are
excited close to the star, where resonantly scattered
photons can convert directly to pairs off the magnetic field.
The loss of charges from the open magnetic field lines
can be balanced by pair creation
within the innermost part of the circuit
if the spin period is shorter than the critical
value given by eq. (\ref{eq:pmax}).
\vskip .2in\null
}
\label{f:gap3}
\end{figure}

A twisted magnetosphere is subject to current-driven instabilities,
which can drive fluctuations in the twist and facilitate its redistribution
across magnetic flux surfaces.  Before considering the time evolution
of the twist in \S \ref{s:eight}, we first consider the influence that
these twist fluctuations would have on the pair density 
in the outer magnetosphere.  We describe a novel state for the open-field
circuit with very strong and efficient particle heating.  Pair creation
is driven by low-energy particles and is concentrated close to the star,
although the heating is distributed over a wide range of radii
(see Fig. \ref{f:gap3}).

Analogous effects are believed to occur in the magnetosphere of one 
star in the double pulsar system J0737$-$3039, which is exposed
to the relativistic wind of the second, faster pulsar.  The dynamic twist
can, in both cases, have important consequences for the particle
energy distribution in the circuit (Thompson \& Blaes 1998; 
Lyutikov \& Thompson 2005).
On some flux tubes, the current that is drawn from the surface of the star will
intermittently be pushed above $J = \rhoGJ c$, which has the effect
of igniting pair creation in order to supply the charges that are 
demanded by the circuit.

Yet higher current densities will result from the
non-linear coupling between torsional Alfv\'en waves propagating 
in opposing directions along the magnetic field, which can create
a high-frequency spectrum of modes.   The necessary condition for
this cascade to occur is that the magnetic field lines are sheared
over a scale (transverse to the magnetic field) that is somewhat
smaller than the radius.  The cascade is not restricted to the closed
magnetic field lines, since an outgoing wave is supplied by reflection
from the surface of the star.

Consider a torsional wave of a moderate amplitude 
\be
\delta B_\phi(r,t) \sim B_\phi 
\ee
which propagates along the dipolar magnetic field $B = \BNS(r/\RNS)^{-3}$.  
Here
\be\label{eq:brms}
B_\phi = \left({J\over \rho_{\rm GJ} c}\right)\,
\left({\Omega r\over c}\right)^{3/2}\,B(r)
\ee
is the static toroidal field at an angle $\theta \sim \theta_{\rm open}$
from the magnetic axis, where
\be
\theta_{\rm open}(r) = \left({\Omega r\over c}\right)^{1/2}
\ee
is the polar angle subtended by the open bundle of magnetic field lines.
In addition, $J$ is the current density that is
drawn from the surface of the star along the open field lines,
$\rhoGJ = {-\Om\cdot\B/2\pi c}$ is the corotation charge density
(Goldreich \& Julian 1969), and $\Om$ is the spin vector of the star.
The wave is decomposed into ingoing and outgoing components,
\be
\delta B_\phi(r,t) 
= \delta B_\phi^-(r+V_wt) + \delta B_\phi^+(r-V_wt)
\ee
which move with a phase speed $V_w$ 
close to the speed of light. We are interested in the
case where the charged particles that support the wave
have a one-dimensional relativistic distribution, which extends up to
some Lorentz factor $\gamma_{\rm max}$.  Then
\be
{V_w\over c} = {\omega\over k_\parallel c} = 
\left(1 + {1\over\sigma}\right)^{1/2}\,\left(1 + 
{k_\perp^2c^2\over 4\gamma_{\rm max}\omega_P^2}\right)^{-1/2}.
\ee
In this expression,
$\omega_P = (4\pi n_\pm e^2/m_e)^{1/2}$ is the non-relativistic
plasma frequency, and $\sigma \simeq B^2/8\pi \gamma_{\rm max} n_\pm m_ec^2$.

The coupling parameter that determines the strength of the interaction
between two colliding Alfv\'en waves is
$k_\perp \xi_\phi = k_\perp \delta B_\phi/k_\parallel B$, where
$\xi_\phi$ is the rotational displacement of the magnetic field lines
and $k_\perp$ and $k_\parallel$ are the components of the wavevector
perpendicular to and parallel to ${\bf B}$.  A cascade develops if 
the shear of the field lines at the forcing scale is stronger than
\be
k_{\perp,0}
\ga  {2\pi\over \theta_{\rm open} r}\,
\left({\delta B_{\phi,0}\over B_\phi}\right)^{-1}.
\ee
Here $\delta B_{\phi\,0}$ is the r.m.s toroidal field at the forcing
scale, which is take to be
\be
{2\pi\over k_\parallel} = L \sim {cP\over 2\pi}.
\ee
This coupling parameter has been
hypothesized to be regulated to $k_\perp\xi_\phi \sim 1$ throughout
the inertial range of the wave spectrum (Goldreich \& Sridhar 1995).
Then a wavepacket undergoes a significant distortion during a single
collision.  If the cascade is characterized by a constant flux of
energy toward higher wavenumbers,
\be\label{eq:ecas}
{dU_{\rm cas}\over dt} \sim 
{\delta B_\phi^2\over 8\pi} k_\parallel c = {\rm const},
\ee
then the Alfv\'en wavepackets become increasingly elongated 
within the inertial range, $k_\perp \propto k_\parallel^{3/2}$.

The fluctuation in the current density associated with the Alfv\'en
waves increases in magnitude toward higher wavenumbers,
\be
\delta J = {c\over 4\pi} k_\perp \delta B_\phi \propto k_\perp^{2/3},
\ee
whereas the energy density in the toroidal field fluctuation decreases,
\be\label{eq:delbk}
{\delta B_\phi^2\over 8\pi} =
{\delta B_{\phi\,0}^2\over 8\pi}\left({k_\parallel L\over 2\pi}\right)^{-1}.
\ee
The current fluctuations at the damping scale are therefore
much larger than the mean current density. 
When the plasma is cold, the damping scale is obtained by
balancing the r.m.s. current density of the wave with the maximum
conduction current that the plasma can supply, 
\be
k_\perp \delta B_\phi \sim 4\pi en_\pm.
\ee
The inner scale of the turbulent spectrum therefore sits at a wavenumber
$k_{\parallel,\rm max} \sim 4\pi en_\pm/B$.

We are, however, interested in the case where the plasma is dilute enough
that the charges become relativistic.  The heated charges
are assumed to move only in the direction parallel to ${\bf B}$.
This assumption is well justified close to the star, where we focus
our attention, due to the rapidity of synchrotron cooling.  In this case,
the particles that support the
fluctuating component of the current must reverse direction:
the current fluctuation due to charges that do not 
reverse direction is negligible, because their velocity undergoes
only a small change, $\delta v \sim c/\gamma^2$.  We deduce that 
the current fluctuation is supported by only a fraction of the
heated charges, with a net density $\delta n_\pm < n_\pm$,
and a characterstic Lorentz factor $\gamma_{\rm damp}$
that is smaller than the mean Lorentz factor 
$\langle\gamma\rangle \sim \delta B_{\phi,0}^2/8\pi n_\pm m_ec^2$.
The associated damping scale is given by
\be\label{eq:kdamp}
k_{\parallel,\rm max} \sim {4\pi e\delta n_\pm\over B}.
\ee

It turns out to be possible to deduce $\delta n_\pm$ and the damping
scale without knowing the distribution function of the charges in 
detail.  At the damping scale, the energy $\Delta E_{\rm damp}$
deposited per charge in one wave period is comparable to the kinetic
energy $\gamma_{\rm damp} m_ec^2$ of the charges that revese direction,
\be\label{eq:edamp}
\Delta E_{\rm damp} \sim {1\over \delta n_\pm}\,
{dU_{\rm cas}\over dt}\,{2\pi \over k_{\parallel\,\rm max} c}
\sim \gamma_{\rm damp} m_ec^2.
\ee
Here
\be\label{eq:ucas}
{dU_{\rm cas}\over dt} \sim \Omega {\delta B_{\phi,0}^2\over 8\pi}.
\ee
(Except for the primary Goldreich-Julian charge flow, 
higher-energy particles come in pairs of opposite signs that
do not gain or lose a significant amount of energy.)
Combining eqs. (\ref{eq:kdamp}) - (\ref{eq:ucas}) then gives the
damping scale
\be\label{eq:kdamp2}
k_{\parallel,\rm max} \sim {2\pi\over \RNS} \left({8\pi^2e\BNS\RNS\over
\gamma_{\rm damp} m_ec^2}\right)^{1/2}\,\left({\Omega \RNS\over c}\right)^2,
\ee
where $\BNS$ is the polar magnetic field strength.
The values of $\gamma_{\rm damp}$ and $k_{\parallel,\rm max}$ can be 
determined independently by the physics of pair creation, 
which leads to a constraint on the spin period (\S \ref{s:cspin}).

A comment about Landau damping is in order here.
We are interested in the regime where the fluctuating component of
the twist is comparable to the mean twist, so that $\sigma \sim 1$.
The Alfv\'en waves then resonate only with the low-energy tail
of the particle distribution, and the Landau damping time is
much longer than the wave period.  Since the lifetime of a
wavepacket is comparable to the wave period in a critically balanced
cascade, the effects of Landau damping can be neglected.
As a result, Landau damping of the Alfv\'en waves is ineffective and the inner 
scale of a turbulent spectrum is located at the
scale where the waves become charge starved.\footnote{See Thompson (2006)
for a description of the conditions in which these
two damping regimes will be found in a cold plasma.}

\subsection{Maximum Spin Period for High Rates of Pair Creation}
\label{s:cspin}

We now consider the range of spin periods and dipolar magnetic fields
that can support rapid pair creation with the circuit model just described.
We focus on the resonant scattering of thermal X-rays by relativistic charges,
which requires only modest particle energies and can therefore be effective
at high particle densities.  The creation of pair-creating gamma rays
by this process is, however, only effective within a relatively small 
distance from the star.

The rate of energy deposition by a turbulent cascade is conveniently
normalized to the corotation particle density
$n_{\rm GJ} = \Omega B/2\pi ec$, so that $(1/n_{\rm GJ})\,
dU_{\rm cas}/dt$ is approximately independent of radius.  If pair
creation is concentrated close to the star, then $n_\pm \propto n_{\rm GJ}
\propto r^{-3}$ on the open magnetic field lines, and the heating rate 
per electron or positron will be approximately constant.
Heating is cut off inside a minimum radius
\be\label{eq:rmin}
R_{\rm min} \sim {1\over k_{\parallel,\rm max}} \ll {c\over\Omega},
\ee
where the energy density in the fluctuating toroidal field 
becomes small, $\delta B_\phi^2/8\pi \propto r^2$.

An upper bound on $R_{\rm min}$ is obtained by requiring that
the magnetic field be strong enough for pair creation.  One therefore
requires a short damping scale and a high pair multiplicity.  
If the spin frequency of the star is too low, then the damping
energy $\Delta E_{\rm damp}$ becomes too small for the heated particles
to produce pair-creating gamma rays.
The energy threshold for pair creation is lower 
for ingoing charges, for which the resonance condition
(\ref{eq:loren}) is most easily satisfied.  We therefore
focus on pair creation by photons that are resonantly scattered
toward the star.

Pairs that are created by this process can be supplied to the outer
magnetosphere only if they reverse direction, that is, only if
their kinetic energy is smaller than $\Delta E_{\rm damp} = \gamma_{\rm damp}
m_ec^2$ (eq. [\ref{eq:edamp}]).
The energy of a resonantly scattered photon can
be related to the kinetic energy $\gamma_{\rm res} m_ec^2$
of the scattering charge via
\be\label{eq:egam}
E_\gamma = f_{\rm recoil}\gamma_{\rm res}m_ec^2 \simeq
\left({B\over \BQ}\right)\gamma_{\rm res}m_ec^2.
\ee
(It will turn out that pair creation is concentrated where
$B<\BQ$, and so we approximate $f_{\rm recoil} \sim B/\BQ$.)
The energy of the gamma ray is approximately divided in two
by pair creation, which leads to the relation
\be\label{eq:gamdamp}
\gamma_{\rm damp} = {B(R_{\rm min})\over 2\BQ} \gamma_{\rm res}.
\ee

Pair creation in the zone $r \sim  R_{\rm min}$
will balance the loss of charges from the circuit if $\gamma_{\rm res}$
(and therefore the energy $\Delta E_{\rm damp}$) exceeds a critical value.
The resonant scattering rate (\ref{eq:Gresin3}) depends 
exponentially on $\gamma_{\rm res}$ when the target photon is 
drawn from the high-energy tail of the black body distribution.
The value of $\gamma_{\rm res}$ that yields a significant rate of
resonant scattering decreases as one moves away from the star, but
its value at a fixed radius depends only logarithmically on
the pair creation rate  through the parameter $\yin$ (eq. [\ref{eq:yine}])
\be\label{eq:gamres}
\gamma_{\rm res} = {B(R_{\rm min})/\BQ\over 2\yin\Th}.
\ee
One requires $\yin \simeq 12$ if $n_\pm/n_{\rm GJ}$ is almost uniform
throughout the open-field circuit.

The values of $R_{\rm min}$ and $\gamma_{\rm damp}$ can now be
determined by requiring that photons of energy (\ref{eq:egam})
able to convert to pairs off the magnetic field.  The corresponding
condition is
\be
E_\gamma {r_{\rm min}\over R_C(r_{\rm min})} > 2m_ec^2,
\ee
where
\be
{R_C(r)\over \RNS} = \eps3^{-1}\left({r\over \RNS}\right)^3
\ee
is the radius of curvature of the field lines, and is
assumed to be dominated by the quadrupole/octopole
component of the magnetic field [see eq. (A9) of Thompson 2008].
One finds,
\be\label{eq:rminb}
{R_{\rm min}\over\RNS} < 3.2\,\eps3^{1/8}B_{\rm NS,15}^{1/4}\,
\left({\yin\over 12}\right)^{-1/8}\,
\left({\kB\Tbb\over 0.5~{\rm keV}}\right)^{-1/8},
\ee
and 
\be
B(r_{\rm min}) > 3.1\times 10^{13}\,\eps3^{-3/8}B_{\rm NS,15}^{1/4}\,
\left({\yin\over 12}\right)^{3/8}\,
\left({\kB\Tbb\over 0.5~{\rm keV}}\right)^{3/8}\qquad{\rm G}.
\ee
The corresponding energy of the pair-creating charges is
\be\label{eq:gavb}
\gamma_{\rm res} > 30\,\eps3^{-3/8}B_{\rm NS,15}^{1/4}\,
\left({\yin\over 12}\right)^{-5/8}\,
\left({\kB\Tbb\over 0.5~{\rm keV}}\right)^{-5/8},
\ee
and the energy of the resultant pairs is
\be\label{eq:gamdampb}
\gamma_{\rm damp}(R_{\rm min}) > 10\,B_{\rm NS,15}^{1/2}\,
\eps3^{-3/4}\,\left({\yin\over 12}\right)^{-1/4}\,
\left({\kB\Tbb\over 0.5~{\rm keV}}\right)^{-1/4}.
\ee

Charges can be supplied self-consistently to the circuit only if
the spin period is  shorter than a critical value, which is obtained
by substituting expressions (\ref{eq:rminb}) and (\ref{eq:gamdampb})
into eq. (\ref{eq:rmin}) and then into eq. (\ref{eq:kdamp2}),
\be\label{eq:pmax}
P < P_{\rm crit} = 7.7\;\eps3^{1/4}\,B_{\rm NS,15}^{1/4}\,R_{\rm NS,6}^{5/4}
\qquad {\rm s}.
\ee
When the spin period is larger than (\ref{eq:pmax}), the cascade deposits
energy too far from the star to deflect ingoing pairs from their creation zone
at $r \sim R_{\rm min}$ back out into the circuit.  This expression has
the attractive feature of depending only weakly on the polar magnetic field
$\BNS$ and the degree of curvature of the magnetic field lines; it is also
independent of $\Tbb$ and $\yin$.  However, the 
normalization does depend on the numerical coefficient (eq. [\ref{eq:rmin}])
relating the minimum radius for pair creation to the damping scale:  one
finds $P_{\rm crit} \propto (k_{\parallel,\rm max} R_{\rm min})^{-1/2}$.
The electric potential fluctuations at the damping scale will, in
practice, have some distribution about $\Delta E_{\rm damp}/e$
(eq. [\ref{eq:edamp}]), with an exponential cutoff at large values.
As a result, the supply of pairs to the outer circuit is possible
if $R_{\rm min}$ is somewhat smaller than the Alfv\'en wavelength
at the damping scale, as is assumed here.

The charges flowing outward toward the open end of the magnetic flux
tube continue to gain energy, because the cascade energy is deposited
broadly throughout the circuit.  The gamma-ray emitting particles at 
$r \sim R_{\rm min}$ have an energy $\gamma_{\rm res} > 
\gamma_{\rm damp}$ (eq. [\ref{eq:gamdamp}]), and therefore have
typically been heated at a larger radius and then partially deflected
back toward the star.  Without prescribing in detail the pair distribution 
function, it is possible to set a lower bound on the multiplicity
in the circuit,
\be
\Mpm = {2\pi e\,n_\pm c\over \Omega B}
   > {2\pi e\,\delta n_\pm c\over \Omega B}.
\ee
Substituting eqs. (\ref{eq:kdamp}) and (\ref{eq:rmin}), one finds
\be
\Mpm  >
  {1\over 2} \left({\Omega R_{\rm min}\over c}\right)^{-1}\nn
  = 4\times 10^3\,\eps3^{-1/8}B_{\rm NS,15}^{-1/4}\,
  \left({P\over 6~{\rm s}}\right)\,
  \left({\kB\Tbb\over 0.5~{\rm keV}}\right)^{1/8}.
\ee
Given a constant outward flux of pairs with a radial velocity comparable
to $c$, the mean energy per charge is
\ba
\langle \gamma(r) \rangle &\sim & {r\over\Mpm n_{\rm GJ}}{dU_{\rm cas}\over dt}
\nn &<&  4\times 10^4\,\eps3^{1/8}B_{\rm NS,15}^{5/4}R_{\rm NS,6}^3
  \left({P\over 6~{\rm s}}\right)^{-3}\,
  \left({\kB\Tbb\over 0.5~{\rm keV}}\right)^{-1/8}
  \left({\delta B_{\phi,0}\over B_\phi}\right)^2\,
   \left({J\over\rhoGJ c}\right)^2
  \left({r\over L}\right).
\ea
Note that $\langle \gamma\rangle > \gamma_{\rm damp}$ at $r > R_{\rm min}$:
most of the heating is at the low-momentum end of the distribution function
$f(P)$.  If $df/dP \sim 0$ at low momentum, then 
$\delta n_\pm/n_\pm \sim \gamma_{\rm damp}/\langle\gamma\rangle$.
But $\delta n_\pm \propto \gamma_{\rm damp}^{-1/2}$, and so we deduce
\be
\gamma_{\rm damp}(r) = \gamma_{\rm damp}(R_{\rm min})\left({r\over
2\pi R_{\rm min}}\right)^{2/3}.
\ee
The damping scale $2\pi/k_{\parallel,\rm max} \propto \delta n_\pm^{-1}$
increases as the ${1\over 3}$ power of the radius.

A competing circuit solution that could lead to such large pair
multiplicities is an outer gap bounded by the surface
$\Om\cdot{\bf B} = 0$ (Cheng, Ho, \& Ruderman 1986;
Cheng \& Zhang 2001).  The voltage across such a gap can
approach the total open-field voltage.  
Seed charges can be created in the gap by collisions between
curvature gamma rays and thermal X-rays from the neutron star surface.
The energetic charges created in such a gap will, however, be beamed
down toward the star, and pair creation will be concentrated 
in a zone where the open field lines are occulted by the star.
Such a voltage structure is therefore not a promising source for
the observed emission of the radio magnetars.  

\section{Variability in the Open-Field Voltage due to Magnetospheric Activity}
\label{s:eight}

The variations in the spindown torque of SGRs and AXPs are remarkable
in their amplitude and longevity:  the spindown rate is observed to 
increase smoothly over a period of months, by up to a factor 
$\sim 4$ (Kaspi et al. 2001; Woods et al. 2002; 
Woods et al. 2007; Tam et al. 2007; Camilo et al. 2007a, 2008).
The strength of the torque variations correlates roughly with the level 
of activity that the magnetar has sustained over the preceding years,
but this correlation is only indirect.  In several cases the torque
increase is observed with a delay of months following an episode of
X-ray burst activity, and can persist long after the intial active period.
The burst emission of
SGRs and AXPs is highly intermittent (Woods \& Thompson 2006):  a nice
example is provided by the AXP 1E 2259$+$586, whose activity as a hard-spectrum
X-ray source was localized within a day, and was followed by an extended
cooling of the thermal X-ray emission (Woods et al. 2004).  
In the most active sources, the magnetar can
remain in a new spindown state for years at a time, with the torque
several times larger than in the preceding state.

Strong variations in spindown torque, extending over a period
of at least several months,
have also been detected in the radio monitoring of the transient AXP 
XTE J1810$-$197 (Camilo et al. 2007b).  This torque
variability persisted long after the initial X-ray activation (in fact,
after the hard-spectrum component of the transient X-ray emission had largely
decayed away).  Even more remarkably, the torque increased by at least
a factor $\sim 3$ compared with older X-ray timing data, making the
relative increase in the spin down rate comparable to that 
seen in the most active SGRs.  

These torque variations have been interpreted in the magnetar model
in terms of a restructuring of the magnetosphere.  Two effects have
been considered which can increase the open magnetic flux, the 
open-field current, and the torque acting on the star.   

\vskip .1in
\noindent 1. --  A persistent static twist 
may be sustained in the closed magnetosphere, driven by the unwinding
of an internal toroidal magnetic field (Thompson et al. 2002;
Beloborodov \& Thompson 2007).  In the process, magnetic helicity is
expelled from the interior of the star.  The twist is stabilized
as long as the outward flux of helicity across the outer boundary of
the magnetosphere is kept small.  

As a dipolar magnetic field is twisted, the poloidal field lines flare 
out slightly.  The large size of the corotating magnetosphere then allows
a significant increase in the open-field current.
In a first approximation, one additional parameter is introduced
into the description of the magnetosphere, namely the twist angle $\Delta\phi$ 
on the field lines that are anchored near the magnetic poles.
In the self-similar construction of Thompson et al. (2002),
the radial index of the magnetic field softens from ${\bf B} \propto r^{-3}$
to ${\bf B} \propto r^{-K}$ with $K < 3$.  The angular width of the
bundle of open field lines therefore increases from $\theta_{\rm open}^2
= \Omega r/c$ to\footnote{This expression is accurate when $K$ is only
slightly smaller than 3;  it neglects any change in the distribution of
flux across the neutron star surface.}
\be
\theta_{\rm open}^2 = \left({\Omega \RNS\over c}\right)^{K-2}\qquad
(K < 3).
\ee
The voltage across the open flux bundle
is proportional to $\theta_{\rm open}^2$.
A twist of one radian forces a reduction in $K$ to 2.85
(see Fig. 2 of Thompson et al. 2002).  At a fixed
rate of spin, the voltage increases by a factor
$(cP/2\pi \RNS)^{3-K} = 4.7\,(P/{\rm 6~s})^{0.15} R_{\rm NS,6}^{-0.15}$,
as does the polar magnetic field that is inferred from the
magnetic dipole model.  

\vskip .1in
\noindent 2. -- A persistent relativistic wind is driven by internal seismic
activity within the star.  This wind combs out the dipolar field lines 
into a radial configuration in the outer magnetosphere (Thompson \& Blaes 1998;
Harding et al. 1999; Thompson et al. 2000).  

The conversion of internal magnetic
and seismic energy to a persistent wind has an uncertain efficiency:
it involves the transfer of energy across magnetic flux surfaces
from an injection zone situated far inside the speed-of-light cylinder.
For example, torsional waves in the outer magnetosphere have a much lower
frequency than do internal elastic modes of the neutron star crust:
the fundamental toroidal mode of the crust
has a period $P_{\rm elastic}\sim 0.02$ s (Strohmayer et al. 1991) and
couples to dipolar field lines with a maximum radius $R_{\rm max}
\sim 0.4\,cP_{\rm elastic}$,  only
$\sim 10^{-2}(P/6~{\rm s})$ of the light cylinder radius.
A wind can therefore be driven by the excitation of waves in the closed
magnetosphere, but the power dissipated on field lines that open out
across the light cylinder may be only a small fraction of the total.

\vskip .1in
Measurements of the torque behavior of magnetars offer ways of
discriminating between these two models.  

A transient twisting up of the magnetic field has
the effect of increasing the open-field voltage and reducing the critical spin
period below expression (\ref{eq:pmax}).  If the detection of the
radio emission from an individual magnetar typically requires a broad 
radio beam, and therefore depends on a high pair density on the open 
field lines (see Thompson 2008), then one can understand why an individual
source would  become visible as a radio pulsar following a period of X-ray
burst activity.   Starting with a value of $P_{\rm crit}$ larger than the 
present spin period, a twisting of the magnetic field can force $P_{\rm crit}$
to a smaller value and ignite a high rate of pair creation on the open field
lines.

The delays that are observed
between burst activity and torque increases can be interpreted
in terms of the resistive evolution of the current within the closed
magnetosphere, which has a timescale of $\sim 1$ yr close to the star
(Beloborodov \& Thompson 2007).  
Their existence is less
clear in the case where the torque change is driven by persistent
seismic activity.
An attractive feature of the twisted dipole model is
that small injections of twist from the
neutron star interior can force large structural changes
in the outer magnetosphere (where the magnetic field energy is relatively
small).  Indeed, the energetic output of
XTE J1810$-$197 during its period of X-ray activity was 
modest by the standards of the Soft Gamma Repeaters;  but the 
amplitude of the subsequent change in spindown torque was as large as is
seen in the most active SGRs.

The measured upper bound on the spin periods of magnetars places additional
strong constraints on how the magnetophere is restructured.  
A persistent wind of particles and Alfv\'en waves drives an exponential
increase in the spin period,
\be
{dP\over dt} \simeq {4\mu_{\rm NS}\over 3I_{\rm NS}}
\left({L_{\rm wind}\over c}\right)^{1/2}\,P,
\ee
where $\mu_{\rm NS} = {1\over 2}B_{\rm NS}\RNS^3$ is the
magnetic moment.\footnote{In the case of a centered dipole, $\mu_{\rm NS}$
can be related to the polar magnetic field $B_{\rm NS}$ and the 
average surface field $\langle B_{\rm NS}\rangle$ via
$\mu_{\rm NS} = {1\over 2}B_{\rm NS}\RNS^3 
= \langle B_{\rm NS}\rangle \RNS^3$.
It is $\langle B_{\rm NS}\rangle$ that is usually quoted in the pulsar
literature.  Hence a polar magnetic field $B_{\rm NS} = 10^{15}$ G, the usual
normalization in this paper, corresponds to $\langle B_{\rm NS}\rangle = 
5\times 10^{14}$ G and a magnetic moment 
$\mu_{\rm NS} = 5\times 10^{32}\,R_{\rm NS,6}^3$ G~cm$^3$.}
This formula applies when the wind luminosity
$L_{\rm wind}$ is much larger than the magnetic dipole luminosity
$L_{\rm MDR} = {2\over 3}\mu_{\rm NS}^3\Omega^4/c^3$.  One finds
\be
{dP/dt|_{\rm wind}\over dP/dt|_{\rm MDR}} = 6.0
{L_{\rm wind,35}^{1/2}\over B_{\rm NS,15}R_{\rm NS,6}}\,
\left({P\over 6~{\rm s}}\right)^2.
\ee

We deduce that a relativistic wind can dominate 
the spindown torque of a magnetar only rarely --- otherwise the spin periods of
magnetars should cover a broad range.  
The efficiency of this 
torque mechanism does not depend in an obvious way on the spin
period and the rate of pair formation the open magnetic field lines.
Alfv\'en waves in the closed magnetosphere are
supported by charges particles of both signs,
that are drawn directly from a light-element surface layer, or
supplied by pair creation near the magnetar surface (as in the 
circuit model of Beloborodov \& Thompson 2007).

\subsection{Timescale for Resistive Evolution of the 
Closed-Field Current and the Open-Field Voltage}\label{s:varia}

Strong variations in radio brightness and pulse shape are detected
from the two radio magnetars on timescales of minutes to hours
(Camilo et al. 2007b, 2008).  These variations have a possible
interpretation in terms of the resistive evolution of currents
in the outer closed magnetosphere, which drives variations
in the open-field voltage.  

Let us first consider the gradual 
redistribution of twist that would result from smooth gradients
in the field-aligned voltage across the poloidal flux surfaces. 
To estimate the timescale in the outer magnetosphere,
we consider the power dissipated on field lines that extend to a
certain maximum radius $R_{\rm max}$.
The self-similar model of a twisted magnetosphere described in
Thompson et al. (2002) has a current density of the form 
\be\label{eq:cur}
{\bf J} = \alpha\,{\bf B};\;\;\;\;\;\;
\alpha =  {c\Delta\phi\over 4\pi \RNS}\,\sin^2\theta_0.
\ee 
Here $\theta_0$ is the polar angle at which a given field line
intersects the star, and $\Delta\phi$ is the net twist angle
between the two magnetic footpoints.  We assume that $\Delta\phi$
is constant in the outer magnetosphere, although in general it
will vary across the poloidal flux surfaces.  Then the toroidal 
magnetic magnetic field is given by 
\be\label{eq:bphi}
B_\phi(r,\theta) = {\Delta\phi\sin^3\theta\over 2r^3}\mu_{\rm NS},
\ee
and the net current flowing across a radius $r > \RNS$ is
\be
I(>r) = \int_0^{\pi/2}\,J_r(r,\theta)\,2\pi r^2
\sin\theta d\theta  = {\Delta\phi\over 8}\,cB_{\rm NS} \RNS
\left({r\over \RNS}\right)^{-2}.
\ee
Given that each closed flux surface develops the same voltage $\Phi_e$,
the dissipation rate is
\be
L_{\rm diss}(>r) = I(>r)\Phi_e.
\ee
The decay time of the twist can be deduced from $L_{\rm diss}$ and the
energy in the toroidal field, which is
\be
E_\phi(>r) \equiv \int_r^\infty r'^2 dr' \int d\Omega {B_\phi^2\over 8\pi}
= {(\Delta\phi)^2\over 210}B_{\rm NS}^2\RNS^3\,\left({r\over \RNS}\right)^{-3}.
\ee
The timescale for the resistive evolution of the twist at radius $r$ is
\be\label{eq:ttwist}
t_{\rm res}(r) \sim {E_\phi(>r)\over L_{\rm diss}(>r)}
 = 0.1\,\Delta\phi\,B_{\rm NS,15}R_{\rm NS,6}^2\,
     \left({|e\Phi|\over {\rm GeV}}\right)^{-1}\,
     \left({P\over 6~{\rm s}}\right)^{-1}\,
     \left({r\over R_{\rm lc}}\right)^{-1}\;\;\;\;{\rm days}.
\ee
Here we have normalized $r$ to the light cylinder radius $R_{\rm lc}
= c/\Omega$.  The field-aligned voltage $\Phi$ is the integral of 
$-E_\parallel = -{\bf E}\cdot\hat B$
along a closed magnetic field line between its two footpoints,\footnote{This
quantity represents the line integral of $(1/c) dA_\parallel/dt$ -- not
the change in the scalar potential, which must vanish given the high
electrical conductivity of the neutron star.}  The value of
$\Phi$ in the outer magnetosphere depends on details
of the return current toward the star:  a gap forming at the surface
of an X-ray bright magnetar has a voltage $e\Phi \sim 1$ GeV
(Thompson 2008), which is similar to the voltage deduced from the 
1-dimensional circuit model of Beloborodov \& Thompson (2007).
At constant $\Phi$, the resistive timescale increases toward the
star.

Our evaluation of $t_{\rm res}$ is 1-2 orders of magnitude longer than
the shortest timescale for radio flux variability observed in the 
magnetars XTE J1810$-$197 and 1E 1547.0$-$5408.  The underlying voltage 
fluctuations may therefore be triggered by current-driven 
instabilities in the outer magnetosphere.  In the next section, we consider
the origin of these instabilities, and how they are regulated by pair
creation.

\subsection{Stability of the Magnetospheric Twist}
\label{s:stable}

A substantial increase in 
the open-field current and spindown torque will result from
a twisting of the magnetosphere only if
the twist is distributed over a broad range of radii. If the twist
were concentrated within a distance $\sim 2\RNS$
of the neutron star, the effect
would be much more limited.  The detection of
strong nonthermal X-ray emission from a magnetar does not necessarily
imply that the source is in its highest torque state.   For example,
the AXP 4U 0142$+$61 
is a copious source of 100 keV X-rays (Kuiper et al.
2006), but its characteristic age is relatively long 
($t_{\rm sd} = P/2\dot P = 70$ kyr); its outer magnetosphere
may therefore be only weakly twisted.

How stable is a magnetosphere with a broadly distributed twist?
A self-similar configuration with a constant
value of the twist angle is not in its minimum-energy
(`Taylor-relaxed') state.  The constant of proportionality $\alpha$ 
between the current density ${\bf J}$ and magnetic flux density 
${\bf B}$ varies with position (eq. [\ref{eq:cur}]).  In particular,
it depends on the polar angle $\theta_0$ at which a given field line
intersects the star, which provides a convenient label for the poloidal
flux surfaces.   In what follows, we allow $\Delta\phi$
to be a function of $\theta_0$.   

Such a configuration with $\alpha$ decreasing
toward the magnetic symmetry axis is susceptible to redistribution of
twist into the outer magnetosphere. The kinetic energy of the charges
that supply the closed-field current is small compared with the
magnetic energy, which means that the magnetosphere is nearly force
free.  The energy of a force-free magnetic field is minimized
at constant $\alpha$ (Woltjer 1958).  When $\Delta\phi$ is constant,
most of the magnetic helicity is stored in the inner magnetosphere:
\be
{\cal H}(r) \;\sim\; \left[\pi r^2 B_P(r)\right]^2\,\Delta\phi(r)
\;\propto\; {\Phi(r)\over r^2}.
\ee
The same is true when $\alpha$ is constant:  in this case $\Delta\phi \propto
1/\sin^2\theta_0 \propto r$, and ${\cal H} \propto r^{-1}$.  The
toroidal field energy is therefore minimized when most of the helicity
is concentrated close to the neutron star.  At the same time,
the redistribution of a small portion of the helicity into the 
outer magnetosphere can drive the twist angle to large values.

To describe the stability properties of the magnetosphere in more detail,
it is useful to define appropriate angular coordinates.  We 
focus on axisymmetric equilibria, for which the magnetic field can
be written as
\be
{\bf B} \;=\; {\bf B}_P + {\bf B}_\phi \;=\; \bnabla\Psi_P\times\bnabla
\left({\phi\over 2\pi}\right) \;+\; \bnabla\Psi_\phi\times\bnabla
\left({\eta\over 2\pi}\right).
\ee
Here $\Psi_P = \mu_{\rm NS}\sin^2\theta/r$ is the flux coordinate
for a dipolar magnetic field,
and $\bnabla\phi = \hat\phi/r\sin\theta$ in spherical coordinates 
($r$,$\theta$,$\phi$).  The toroidal magnetic field corresponding
to eq. (\ref{eq:cur}) is given by eq. (\ref{eq:bphi}).
The toroidal flux threading a poloidal flux surface of cross-section
$S$ that is anchored at angle $\theta_0$ is
\be\label{eq:phint}
\Psi_\phi(\theta_0) = \int_S r dr d\theta B_\phi(r,\theta).
\ee
The surface integral
(\ref{eq:phint}) can be transformed into an integral over 
$\theta$ and $\theta_0$ using the relation $r = \RNS\sin^2\theta/
\sin^2\theta_0$,
\be
\Psi_\phi(\theta_0) = {\Delta\phi\mu_{\rm NS}\over \RNS}
\int_{\theta_0}^{\pi/2} \sin\theta_0'\cos\theta_0'\, d\theta_0'
\int_{\theta_0'}^{\pi-\theta_0'}\sin\theta\,d\theta.
\ee 
The distribution of toroidal flux with surface polar angle is
\be
{d\Psi_\phi\over d\sin\theta_0} = {2\Delta\phi\,\mu_{\rm NS}\over \RNS}
\sin\theta_0\cos\theta_0,
\ee
and the dual angular variable is
\be
{\eta\over 2\pi} = {1\over 2}\left(1-{\cos\theta\over \cos\theta_0}\right).
\ee

The strength of the winding in the magnetosphere can be described
in terms of the relative periodicities in the $\phi$ and $\eta$ directions,
\be\label{eq:safety}
{1\over q(\theta_0)} \equiv {d\phi\over d\eta} = {B_\phi\over B_\theta\,
\sin\theta}\,
 \left({d\eta\over d\theta}\right)^{-1}  =
{\Delta\phi(\theta_0)\over 2\pi}\cos\theta_0.
\ee
A twisted magnetic field is subject to a variety of current-driven
instabilities.  Ideal MHD instabilities set in at strong values of
the twist: the kink instability is present when the `safety factor'
$q \leq 1$ in cylindrical geometry (e.g. Wesson 2004).  On the other
hand, the flaring of the dipolar magnetic field lines has a strong effect on
the spindown torque even when $\Delta\phi \simeq 0.5-1$, which corresponds to
$q \simeq 5-10$.  The hard X-ray emission of magnetars can be powered
by even weaker twists in the inner magnetosphere:  only $\Delta\phi \sim 0.1$
is required (Thompson \& Beloborodov 2005, Beloborodov \& Thompson 2007),
corresponding to $q \sim 10^2$. 

The redistribution of twist by tearing instabilities within the closed 
magnetosphere of a neutron star must therefore be considered.  
Resistive kink instabilities are known to trigger a redistribution of current
across magnetic flux surfaces in tokamak plasmas.  They can be excited
at lower values of the twist, and are sensitive to the current 
distribution within the tokamak torus.\footnote{Our notation
here is the inverse of that in the tokamak literature:  the unstable
component of the twist is in the azimuthal direction $\phi$, whereas
in a toroidal tokamak plasma it is in the $\theta$ direction.  The poloidal
magnetic field of a neutron star is supported by currents embedded in
the star, and the external toroidal field by currents flowing through the
magnetospheric plasma.  In the case of a tokamak, the
toroidal magnetic field is supported by external currents that
loop around the long axis of the torus, and the poloidal 
field ($B_\theta$) is supported by plasma currents.}  

\begin{figure}
\epsscale{1.1}
 \plottwo{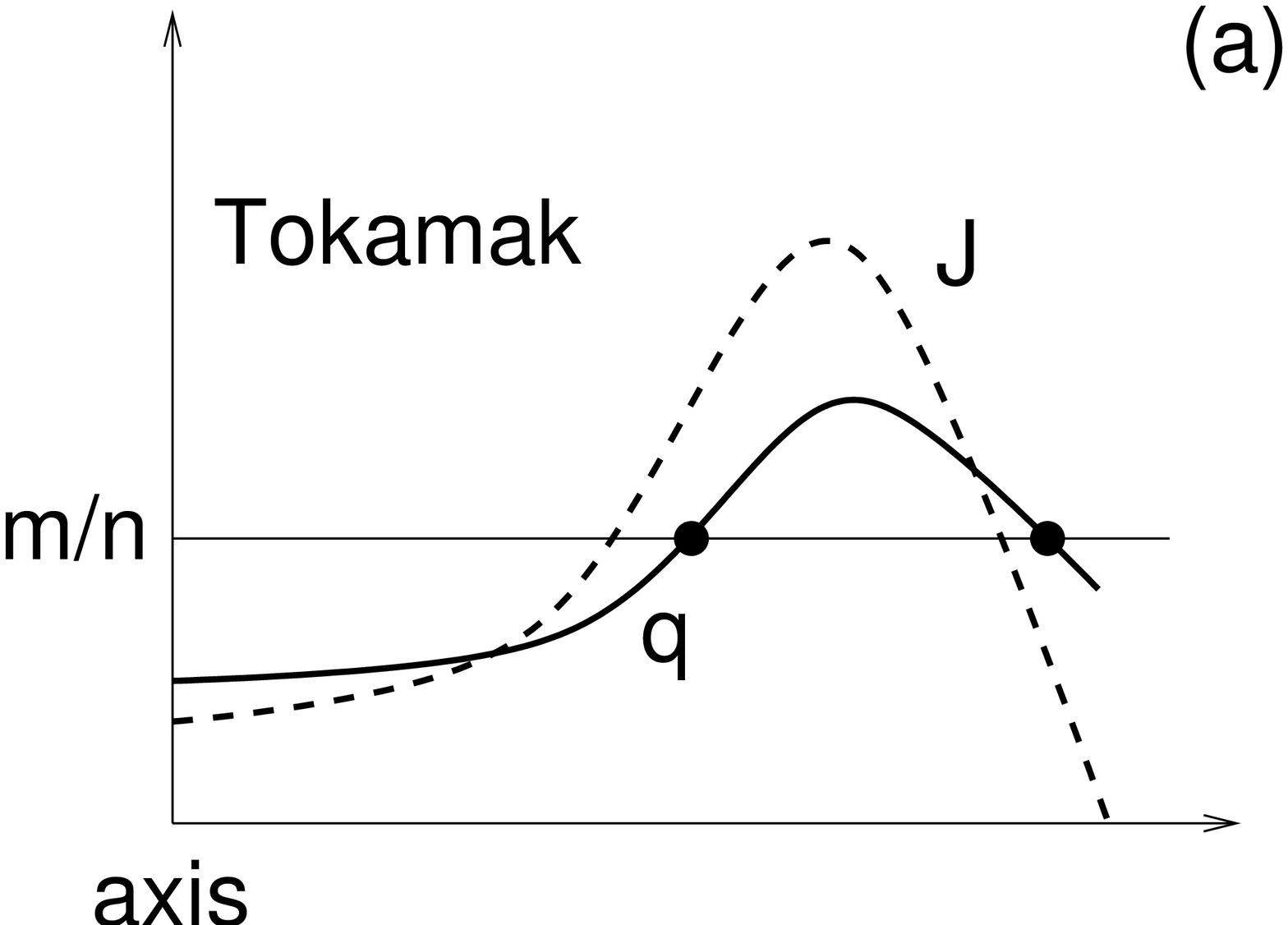}{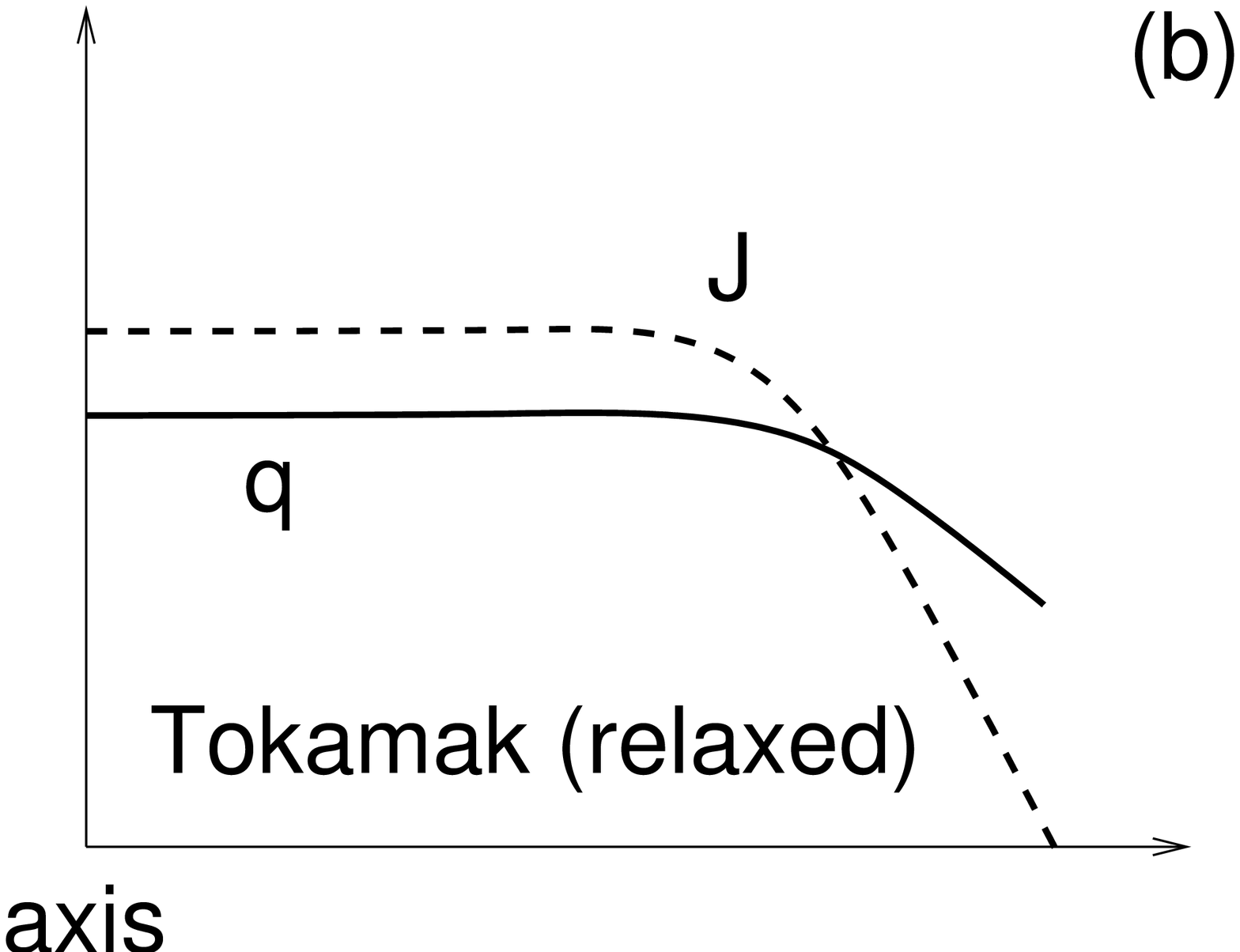}
\vskip .3in
\epsscale{0.55}
\plotone{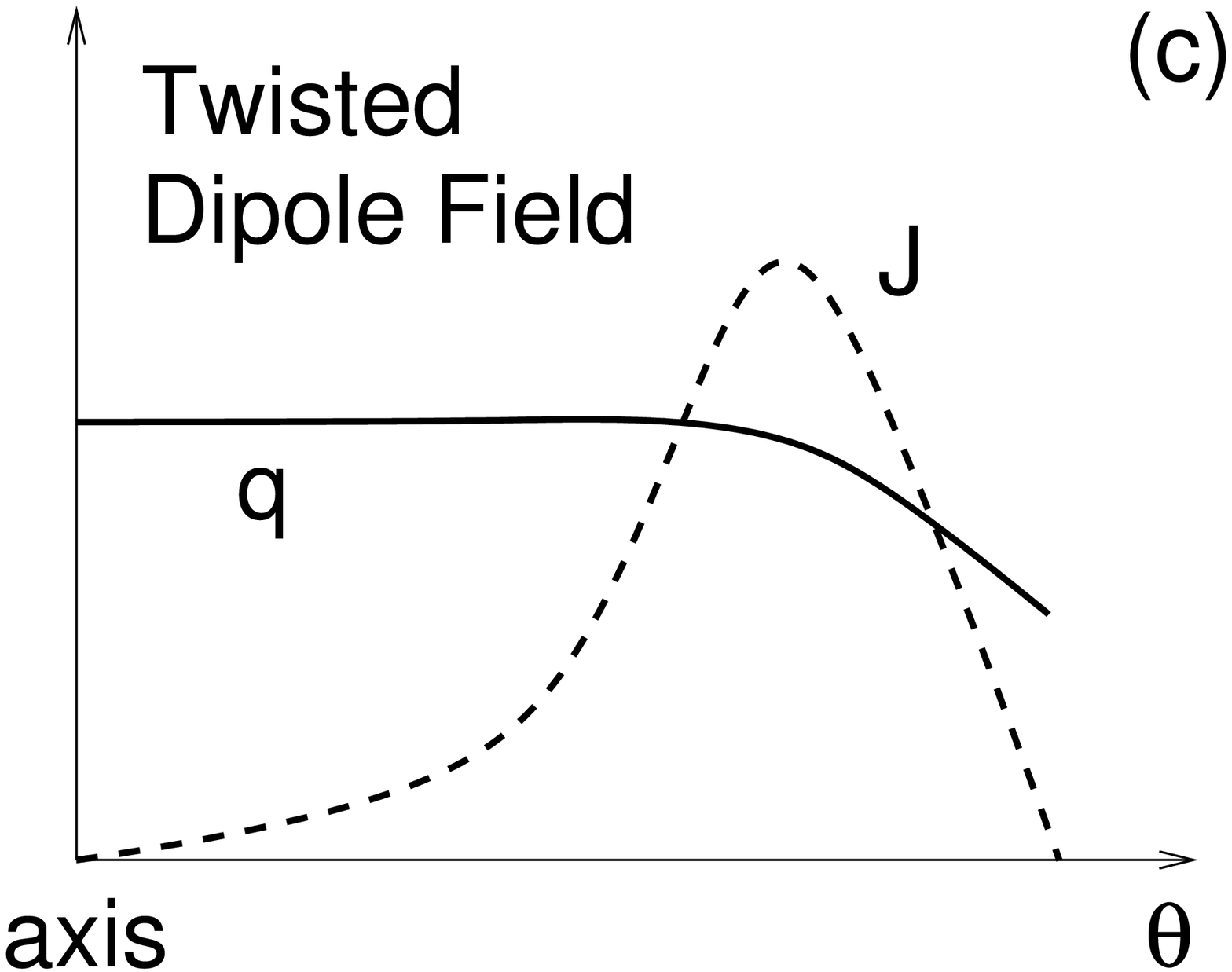}
\caption{Cross section of the current density $J$ and safety factor $q$
in a plane perpendicular to the long axis of a tokamak torus 
(panels a,b) and to the symmetry axis of a twisted dipole magnetic
field (panel c).  In a tokamak, $q$ must have a minimum
near the axis if $J$ does.  A given mode then can interact with
two rational surfaces, which accentuates the growth rate of 
current sheets at the rational surfaces.  The net effect is a rapid
flattening
of the current and safety factor profiles.  By contrast, the $q$ profile can
remain quite flat near the symmetry axis in a neutron star
magnetosphere, even through $J$ has a minimum on the axis.
\vskip .2in\null}
\label{f:current}
\end{figure}

Here there is an
important distinction between a tokamak plasma and the twisted
magnetosphere of a neutron star.  The toroidal current in a
tokamak plasma torus is observed to relax rapidly to a constant-$\alpha$
state if the current is concentrated initially near the outer boundary
of the torus.   As a result, the profiles of the current and the
safety factor $q$ both become much flatter in near the long axis
of the torus (Fig. \ref{f:current}a,b).  
The outer magnetosphere of a neutron star
can, by contrast, sustain a current distribution in which
the $q$-profile is flat across the poloidal flux surfaces ($\Delta\phi$ is
constant) even though the current density decreases toward
the magnetic axis, $J \propto \theta^2$ (Fig. \ref{f:current}c).

This difference in the shape of the $q$-profiles in unrelaxed
tokamak and neutron star plasmas has interesting implications for
the excitation of resistive kink instabilities.  
The tearing modes which facilitate the redistribution
of twist across the (toroidal) flux surfaces are concentrated on
rational surfaces where 
\be
({\bf B}\cdot\bnabla)\delta{\bf J} = {iB_\phi\over R}
\left({m\over q} - n\right)\delta{\bf J} = 0.
\ee
Here $\delta {\bf J} \propto e^{i(m\theta-n\phi)}$ 
is the current perturbation and $R$ is the large radius of the plasma torus.  
The analogous form for the current
perturbation in a neutron star magnetosphere is
\be
\delta{\bf J} \propto \exp\left[im\phi - 
i\pi n\left(1-{\cos\theta\over\cos\theta_0}\right)\right],
\ee
where the roles of the $\theta$ and $\phi$ angular variables are reversed.
It is straightforward to check that
\be
({\bf B}\cdot\bnabla)\delta{\bf J} = {iB_\theta\over r}
\left( {\Delta\phi\over 2\pi}m - n\right)\,\delta{\bf J},
\ee
where we have set $\cos\theta_0\rightarrow 1$ for the most
extended dipole field lines.

When the current distribution across the long axis of a tokamak torus has
a localized maximum, then so does the distribution of $q$.  As a result,
a single resonance $q = m/n$ can appear at two separate surfaces
within the torus (Fig. \ref{f:current}).  
These double tearing modes appear to be important in the process of 
Taylor relaxation (e.g. Stix 1976).  They can be avoided in the
twisted magnetosphere of a neutron star if the twist has the same sign
on the open and closed magnetic field lines in both magnetic hemispheres.
The presence of a highly conducting boundary also tends
to suppress tearing modes at rational surfaces near the outer boundary
of tokamak plasmas (e.g. Wesson 2004). 

The spin of the star induces a twist on the open magnetic field lines
with the same sign as the twist that is sustained by the outward
Goldreich-Julian current.  A reversal in the sign
of the twist on the open field lines, if sustained over many rotation
periods, therefore demands a reversal in the sign of the conduction
current.  This is possible only if a polarizable plasma is present on 
the open  field lines:  the space charge density on the open magnetic field
lines just above the surface of the neutron star must be the same
as $\rho_{\rm GJ}$.  On this basis, Thompson et al. (2002) conjectured
that the rate of reconnection near the magnetospheric boundary would
be sensitive to the pair production rate in the open-field accelerator.

The stability of the twist in the outer magnetosphere depends on whether
the safey factor $q$ (eq. [\ref{eq:safety}]) can maintain a flat profile
across the open-field boundary.  In the standard
pulsar model, the open-field current in each hemisphere is compensated
by a return current of the same magnitude that flows along a narrow
annulus of closed field lines just inside the magnetospheric boundary.  
If the open-field current had the
opposite sign to the Goldreich-Julian current in one hemisphere, then
the toroidal field would go to zero on the flux surface bounding the
return current sheath.  This would have the effect of allowing
double rational surfaces to form, in close analogy with a Tokamak
plasma (Fig. \ref{f:current}).   By contrast, when the open-field current
is able to reverse sign, then
the $q$-profile can remain flat and double rational surfaces do not
form.

\section{Clustering of Spin Periods in the Magnetar Population and\\
the Sudden Brightening of Radio Magnetars}\label{s:seven}

X-ray bright magnetars\footnote{Observed with persistent luminosities of 
$10^{35}$ erg s$^{-1}$ or above over a period of years or decades.
See the McGill Magnetar web page for an
up-to-date list of magnetar properties; 
{\tt http://www.physics.mcgill.ca/$\sim$pulsar/magnetar/main.html}.}
range in spin periods between 0.33 and
11.8 seconds.  The upper cutoff to the spin distribution
is statistically significant (Psaltis \& Miller 2002). 
The distribution is also quite strongly peaked:  removing two sources
(1E 1547.0$-$5408, $P= 2$ s, and PSR J1846$-$0258, $P = 0.33$ s), 
the remaining spin periods lie between $5.2$ and $11.8$ s.  
If the stars underwent simple magnetic dipole spindown, then
a range of 2.3 (6) in spin period would correspond to a range of
$\sim 5$ ($\sim 30$) in age.  
The measured characteristic ages actually have a much wider
distribution, from several hundred years in the most active SGRs to
$2\times 10^5$ yrs in the AXP 1E 2259$+$586.
The observation of strong fluctuations in spindown torque in individual
sources shows that part of the dispersion in
characteristic ages is due to variations in magnetospheric
structure.  This measured torque variability cannot be related a narrowing
of the spin period distribution unless the underlying mechanism 
is sensitive to the rate of pair creation in the outer magnetosphere.  

We are interested in the case where the rate of injection
of toroidal field energy into the inner magnetosphere (when averaged over
the episodes of X-ray activity) is much higher
than the spindown power.  The net twist on the magnetic field lines that
thread the closed magnetosphere and the interior of the star can be
described by the magnetic helicity ${\cal H} = \int {\bf A}\cdot\B\,dV$,
where ${\bf A}$ is the vector potential.  
The toroidal magnetic field in the closed magnetosphere is
susceptible to reconnection with the external magnetic field.
We explained in \S \ref{s:stable} how this process could be impeded
by a high rate of pair creation on the open magnetic field lines
(see also Thompson et al. 2002).   

A slow loss of helicity across
the outer boundary of the closed magnetosphere can force large-amplitude
fluctuations in the twist on the open magnetic field lines, and
reverse its sign in one hemisphere.   The required rate of loss is
\be\label{eq:hloss}
{d{\cal H}\over dt}\biggr|_{\rm lc} \simeq
-\left[\pi R_{\rm lc}^2 B_P(R_{\rm lc})\right]^2\,\Omega,
\ee
where $R_{\rm lc} = c P/2\pi = 3\times 10^5\,(P/6~{\rm s})$ km.
The corresponding damping time of the twist $\Delta\phi$ close
to the star is therefore
\be
{\Delta\phi(r)\over |d\Delta\phi(r)/dt|_{\rm lc}}
\simeq 30\,{\Delta\phi\over R_{\rm NS,6}}
\left({P\over 6~{\rm s}}\right)^3 
\qquad {\rm yr}.
\ee
In the case of a magnetar, the damping time in the inner
magnetosphere is comparable to the resistive decay time of several
years (eq. [\ref{eq:ttwist}]).  

We have shown in \S \ref{s:six} that a very high pair density can be sustained
in the outer magnetosphere if a dynamic twist is excited and then
damped through a turbulent cascade to high spatial 
wavenumbers.   This process depends on a high rate of particle heating 
close to the star, where the rate of resonant scattering is high, and 
the resonantly scattered photons are energetic enough to convert to pairs.  
The critical spin period $P_{\rm crit}$ (eq. [\ref{eq:pmax}]) for such a 
gas to be sustained on the open magnetic field lines is several seconds,
and depends weakly on the polar magnetic field strength and the 
curvature of the open magnetic field lines.

\subsection{Metastable Twisted Magnetosphere and Torque Decay}

We are led to conjecture that the outer magnetosphere can enter a metastable
state in which the twist is lost across the magnetospheric boundary at
a controlled rate, just high enough to excite large fluctuations in the
twist over a single rotation of the star.  
This metastable state depends on the presence of a dense pair
gas in the outer magnetosphere, which can limit the rate of reconnection
of the closed magnetosphere with the open-field bundle.  
The detection of hard X-ray emission at a level
$L_X \simeq 10^{35}$ ergs s$^{-1}$
from AXPs with little known X-ray burst activity (Kuiper et al. 2004, 2006)
leads to the remarkable inference that
magnetic helicity can be stored in the magnetosphere for several
years.  Hard X-ray emission of a comparable luminosity
has been detected from SGR 1900$+$14 in
a relatively quiescent state, nearly a decade after the last major
X-ray burst activity (G\"otz et al. 2006b).  Very bright emission
($\sim 10^{36}$ ergs s$^{-1}$) was detected from SGR 1806$-$20 by
Integral within two years of its major 2004 outburst (Mereghetti et al. 
2005).  

The observed peak in the distribution
of magnetar spins can then be explained if most of the spindown torque
is accumulated in this metastable, twisted state, and not when the
magnetosphere is more nearly dipolar.  Once the spin period 
of the star grows beyond the value (\ref{eq:pmax}), the magnetosphere
cannot sustain the inflated, twisted state.  Of course, there is some
hysteresis in thie process, because a high rate of pair creation
can be reignited by the process of twisting (\S \ref{s:eight}), but
this only introduces a spread of a factor $\sim 2$ in the value
of $P_{\rm crit}$.  We also emphasize that this does not exclude an
important contribution of a particle wind to the torque when the
outer magnetosphere is nearly dipolar (Spitkovsky 2006).

We must therefore examine whether most of the spindown can realistically
occur when the magnetosphere is twisted.  The duty cycle of the
magnetospheric twist can be determined by combining the magnetospheric
dissipation rate (which can be measured in the hard X-ray band)
with net twist of the internal magnetic field.  The internal field
cannot be directly measured, but the energetics of the giant flares
of the SGRs point to the presence of internal magnetic fields 
approaching $10^{16}$ G (e.g. Hurley et al. 2005).

We suppose that, on multiple occasions,
a certain amount of helicity is injected into the magnetosphere
due to the release of sub-surface stresses, and is stored there.  The
decay time of the twist can be deduced from eq. (\ref{eq:ttwist}),
by substituting explicitly for the magnetospheric dissipation rate
$L_{\rm diss}$,
\ba\label{eq:tdecay}
t_{\rm decay} &=& {E_\phi\over L_{\rm diss}} = {32\over 105}\,
{L_{\rm diss} \RNS\over c^2\Phi_e^2}\nn
&=& 0.1\,L_{\rm diss,35}R_{\rm NS,6}\,
\left({\Phi_e\over 10^9~{\rm V}}\right)^{-2}\;\;\;\;{\rm yr}.
\ea
If the current is concentrated on poloidal field lines that are
anchored within an angle $\theta < \theta_c$ of the magnetic axis,
then it is easy to show that $L_{\rm diss} \propto \theta_c^4$
and $t_{\rm decay} \propto \theta_c^2$.  

The net duration $t_{\rm twist}$ of the twisted state in the magnetosphere is
determined by the winding angle $\Delta\phi_i$ of the internal magnetic field.
Given that a toroidal magnetic field is present to a depth
$\Delta R$ below the stellar surface, one has 
$\Delta\phi_i \sim (\Delta R/\RNS)\,B_{\phi,i}/B_{\rm NS}$ and
\ba
t_{\rm twist} &\simeq&
 {\Delta\phi_i\over \Delta\phi}\,t_{\rm decay}
= {4\over 105}{B_{\phi,i}\,\RNS\,\Delta R\over \Phi_e c}\nn
&=& 120\,R_{\rm NS,6}^2\,\left({B_{\phi,i}\over 10^{16}~{\rm G}}\right)\,
\left({\Phi_e\over 10^9~{\rm V}}\right)^{-1}\,
\left({\Delta R\over \RNS}\right)\;\;\;\;{\rm yr}.
\ea

The spindown rate of the magnetar is accelerated when its magnetosphere
is twisted.  We define 
\be
(\dot\nu)_{\Delta\phi} = N_b \dot\nu_0,
\ee
where $\dot\nu_0$ is the frequency derivative in the absence of twisting.
The acceleration factor is quite large, $N_b \sim 10$,
if the twist angle is as large as 1 radian (Thompson et al. 2002),
but observations suggest $N_b \la 4$.  
Given that $t_{\rm twist}$ is smaller than the total age\footnote{We
neglect any departures from a simple magnetic dipole torque model,
except for those due to twisting.  Thompson et al. (2002)
 found that the braking index 
$n = \ddot\nu \nu/(\dot\nu)^2$ is slightly smaller than
 the dipole value ($n=3$) 
in the twisted state, but this feature is not important for present purposes.}
$t_{\rm sd,0}$ 
of the magnetar, the relative frequency change due to spindown in the
twisted and untwisted states is
\be
{(\Delta \nu)_{\Delta\phi}\over (\Delta\nu)_0}
= N_b\,{t_{\rm twist}\over t_{\rm sd,0}}.
\ee

Consider an SGR with an observed spindown age $t_{\rm sd,0}/N_b = 200$
yrs, which is the minimal value obtained from the spin monitoring of
SGR 1806-20 (Woods et al. 2007).  Most of the 
spindown will occur in the twisted state, 
$(\Delta\nu)_{\Delta\phi} > (\Delta\nu)_0$, if
\be
\left({B_{\phi,i}\over 10^{16}~{\rm G}}\right)\,
\left({\Phi_e\over 10^9~{\rm V}}\right)^{-1}\,
\left({\Delta R\over \RNS}\right) > 1.7,
\ee
and if the loss of helicity due to non-axisymmetric instabilities
in the magnetosphere can be neglected in comparison with
ohmic damping in situ.
If the voltage is as small as $\sim 10^8$ V, then
twisting of the magnetosphere can continue to enhance the time-averaged
torque as long as the spindown age is shorter than $\sim 10^4$ yrs.

\subsection{Torque Decay due to Dipole-Spin Axis Alignment?}

In this picture, the spin evolution of magnetars is driven by
departures from a dipole structure in the closed magnetosphere.
The limiting spin period for magnetars is then not directly tied to
the radio death line of pulsars.  We should, nonetheless, consider
seriously the possibility that an active pair accelerator has been
entirely quenched on the open magnetic field lines of many magnetars.
X-ray pulse measurements of magnetars would then provide valuable information
about the electrodynamic properties of pulsars on both sides of the
radio death line.

This intriguing possibility encounters two basic difficulties. 
First, pair creation by the resonant scattering of thermal X-rays
is very effective near the X-ray bright surface of a magnetar,
and a pair discharge can be maintained on the open magnetic field lines
at spin periods significantly longer than the observed
range of magnetars (e.g. Medin \& Lai 2007).
Second, a reduction in the open-field particle luminosity would result
in a large reduction in torque only if the magnetic dipole and rotation
axes were nearly aligned, so that the magnetic dipole luminosity
in vacuum is much smaller than the particle luminosity in the active
state (e.g. Contopoulos \& Spitkovsky 2006).  

Observations of PSR
B1931$+$24, which undergoes radio nulls of a month-long duration,
show the torque is reduced by a factor $\sim {2\over 3}$ when the radio
emission is off (Rea et al. 2006).  If $\bmu$ and $\Om$
are inclined by an angle $\chi_{1931}$ in this object, then
then one deduces that the ratio
of magnetic dipole luminosity to particle luminosity is, more generally,
\be
{L_{\rm MDR}\over L_{\rm particle}} \simeq 2{\sin^2\chi\over\sin^2\chi_{1931}}
\ee
in a radio-bright state.  The characteristic age of the AXP 1E 2259$+$586
is $\sim 2\times 10^5$ yrs, some 20 times longer than the estimated
age of the surrounding supernova remnant CTB 109 (Sasaki et al. 2004). 
To explain
a reduction of a factor $\sim 20$ in torque following the quenching
of radio emission, one would require that $\sin^2\chi < \sin^2\chi_{1931}/40$.

There are strong arguments against the alignment of
$\bmu$ and $\Om$ in most magnetars.  
The rotational bulge of a slowly rotating magnetar
is much smaller in amplitude than the variation in the lengths of the
principal axes due to
the internal magnetic field.  When the internal field is predominantly 
toroidal, the star becomes prolate in shape and its two largest
principal moments of inertia lie in the plane perpendicular to the
magnetic symmetry axis.  The star reaches a state of minimum
rotational energy (at fixed spin angular momentum)
when $\Om$ lies in this plane (e.g.
Cutler 2002).  If the external magnetic moment were aligned with
the symmetry axis of the internal field, then there would be no
reason to expect alignment between $\bmu$ and $\Om$.

A near alignment between $\bmu$ and $\Om$ would be easier
if the magnetic moment were distributed randomly with 
respect to the axis of the internal toroidal field.  But in that
case, the probability of finding alignment to the degree of precision
required for 1E 2259$+$586 is only ${1\over 2}\chi^2 \sim 0.01$.

\section{Summary}\label{s:summary}

\noindent 1. -- A gamma ray that is created below the threshold energy
for pair conversion can adiabatically convert to bound positronium
(Wunner et al. 1985; Shabad \& Usov 1986).  We show that such a positronium
atom is rapidly dissociated in the presence of an
intense flux of thermal X-rays from the magnetar surface,
through the resonant absorption of an X-ray by one of the two charges.  
Positronium is also be photodissociated by infrared radiation if
the surface luminosity at $\sim 0.1$ eV exceeds $\sim 10^{29}$ ergs s$^{-1}$.
This minimum luminosity
is only $\sim 10^{-2}-10^{-3}$ of the observed infrared output of the
AXPs, but some $\sim 10^6$ times the flux expected from the Rayleigh-Jeans
tail of the X-ray blackbody.

\vskip .1in
\noindent 2. --  We have calculated the cross section for
the reaction $\gamma + \gamma \rightarrow e^+ + e^-$ in the presence
of an ultrastrong magnetic field, using electron wavefunctions that
are eigenstates of the electron spin.  We find a resonance in the
cross section when the gamma ray is, by itself, just below the threshold
energy for pair creation.  Away from this resonance, the value 
of the cross section differs slightly from previous results.  

\vskip .1in
\noindent 3. -- We have considered the generation of $e^\pm$
pairs when the magnetic field lines are strongly turbulent, and have
found a self-consistent state with a very high pair density and
a low mean energy per particle.   In this
situation, a high-frequency spectrum of Alfv\'en waves is created via
a turbulent cascade.  The current density diverges in this cascade.
There is a critical wave frequency above which the 
plasma is not able to support the current, and the energy
of the waves is transferred to the longitudinal motion of the charges.
When the pair density is high, particles are heated close to the star,
where resonant scattering is effective.  There is a critical spin period
of $\sim 5-7$ seconds, below which the pair creation rate in
this inner zone is high enough to compensate the loss of pairs across
the light cylinder and into the pulsar wind.  
This result does not depend on the details of the particle distribution
function within the circuit.   

\vskip .1in
\noindent 4. -- The timescale
for the ohmic evolution of the twist in the outer magnetosphere is
relatively short ($\sim 0.1$ day).  The fast variability that
is observed in the brightness of the radio magnetars therefore has a 
plausible explanation in terms of current-driven instabilities on 
the closed magnetic field lines.  

\vskip .1in
\noindent 5. --  We have described in some detail a conjecture 
that the outer magnetosphere
can remain strongly twisted only when the pair density is very high
on the open magnetic field lines.  In this possible metastable 
state, the loss of magnetic helicity across
the magnetospheric boundary is just high enough to sustain large
fluctuations in the twist, and high rates of pair creation.  
Part of the motivation comes from the
consideration of the growth of current-driven instabilities in the
outer magnetosphere, which are strongly enhanced if 
the sign of the current on the open field lines matches that flowing
into the polar regions of the rotationally-driven wind.  

\vskip .1in
\noindent 6. --  The cumulative torque that can be exerted on the
spin of  a magnetar in the active magnetospheric phase has been
considered, and it has been shown that it can 
exceed the cumulative magnetic dipole torque.  For this reason, the
observed clustering of magnetar spins may be connected to a loss
of stability of a twisted magnetosphere at a characteristic spin
period of several seconds

\vskip .1in
As is discussed in further detail in Thompson (2008), the dynamic model
of the outer magnetosphere investigated here has a promising application
to the radio magnetars, in particular to their hard spectra, 
broad pulses, and relatively high luminosities (in comparison with
the spindown luminosity).   The consistency of
this pair-rich circuit solution does not, by itself, imply a radio
death line for magnetars -- a weaker pair discharge
appears possible at significantly longer spin periods (e.g. Medin \& Lai 2007).
However, radio emission from the inner magnetosphere would be much
harder to detect, due to the effects of beaming.

\acknowledgments
This work was supported by the NSERC of Canada.  I am indebted to
Andrei Beloborodov for explaining some of the literature on
positronium creation to me.  His suggestion that
a hybrid gamma-ray/electron-positron pair might be able to
photodissociate led to the calculation in \S \ref{s:case2}.
I also thank Cornell University for its hospitality when part
of this work was completed.

\begin{appendix}

\section{Calculation of Cross Section for $\gamma +\gamma \rightarrow
{\sl e}^+ + {\sl e}^-$}

\begin{figure}
\epsscale{0.8}
 \plotone{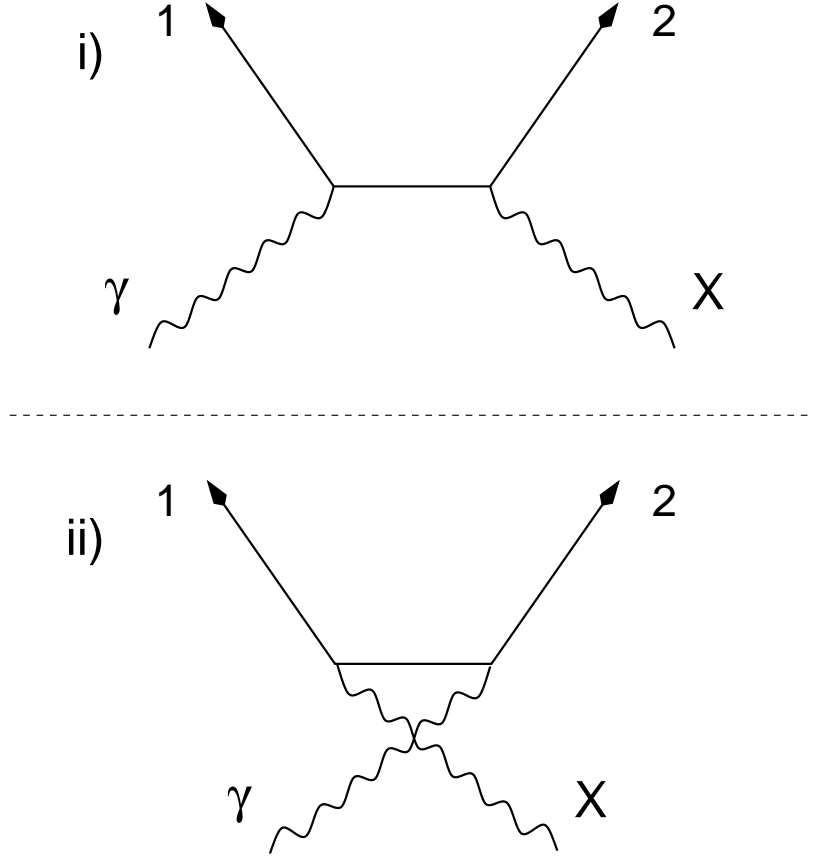}
\caption{Diagrams contributing to the calculation of the 
cross section for $\gamma + \gamma \rightarrow e^+ + e^-$. 
In case 1 (head on collision between X-ray and gamma-ray),
the outgoing charges 1 and 2 are both in their lowest Landau
state $n=0$.  In case 2 (gamma ray near threshold for pair
single-photon pair creation), the charge that shares a
vertex with the gamma-ray ends up in Landau state $n=0$,
and the charge which shares a vertex with the X-ray ends
up in the state $n=1$.  When the interaction between the
two photons is resonant, there is no interference between
the two channels.
\vskip .2in\null
}
\label{f:diagram}
\end{figure}

We summarize our calculation of the creation of an electron-positron
pair in an ultrastrong magnetic field, via the collision of two photons.
The basic strategy is to calculate the tree level matrix elements illustrated
in Fig. \ref{f:diagram}, using magnetized 
wavefunctions for the outgoing electron and positron and the virtual
electron.  In this section, we use naturalized units in which
$\hbar = c = 1$.

We focus on two special cases:  

\vskip .1in
\noindent 1. -- The two photons collide head on, each with a center
of mass energy $\omega_1\ = \omega_2 > m_e$.  Their
momenta are aligned with the background magnetic field ${\bf B}$.  This 
choice of ingoing momenta corresponds approximately to the case
where an energetic gamma-ray is created by resonant scattering of a thermal
X-ray by a relativistic charge, in a magnetic field $B < 4\BQ$. 
Then the gamma ray is below threshold for single-photon pair creation
near the point of creation, $\omega_1\sin\theta_{kB,1} < 2m_e$, 
but it can create a pair by colliding with
a thermal X-ray.  In the frame where the net momentum parallel to 
${\bf B}$ vanishes, $k_{1,z} + k_{2,z} = 0$, the target
X-ray propagates nearly parallel to ${\bf B}$, and the gamma ray 
propagates at a modest angle to the magnetic field.

\vskip .1in
\noindent 2. -- One of the photons is close to the threshold for
single-photon pair creation, 
\be
\omega_1\sin\theta_{kB,1}  = 2m_e - \Delta E.
\ee
We take this photon to propagate perpendicular to ${\bf B}$,
$\theta_{kB,1} = \pi/2$, and the target photon parallel to ${\bf B}$
(Fig. \ref{f:gXcol}).  
The second photon need not have a very large energy to permit
the creation of an electron and positron.

\vskip .1in
A calculation of the cross section for 
$\gamma + \gamma \rightarrow e^+ + e^-$
in an ultra-strong magnetic field was published by Koslenkov
\& Mitrofanov (1986) using the Johnson \& Lippman (1949) electron/positron
wave functions.  These wave functions are not, however,
eigenstates of the electron spin.  We therefore use an alternative 
form derived by Sokolov \& Ternov  (1968) and Melrose \& Parle (1983).

The solution to the Dirac equation
\be
\left[\gamma^\mu\left(i{\partial\over\partial x^\mu} - eA_\mu\right) 
- m\right]\psi = 0
\ee
is labeled by the component of the spin parallel to the magnetic field,
${\bf\sigma}\cdot\hat B = {1\over 2}\sigma = \pm{1\over 2}$,
the longitudinal momentum $P = {\bf P}\cdot\hat B$, and the Landau level $n$:
\be
\left[\psi_-^{(\sigma)}({\bf x},t)\right]_{P,n,a}
 = e^{iP\cdot x}u^{(\sigma)}_{n,a}({\bf x}_\perp),
\ee
for the electron (positive-energy) states,
and
\be
\left[\psi_+^{(\sigma)}({\bf x},t)\right]_{P,n,a} = 
e^{-iP\cdot x}v_{n,a}^{(\sigma)}({\bf x}_\perp)
\ee
for the positron (negative-energy) states.  The dispersion relation
is given by
\be\label{eq:disrele}
E^2 = P^2 + E_{0n}^2 = P^2 + m_e^2 + 2n|e|B,
\ee
where $P$ denotes the component of the momentum parallel to ${\bf B}$.
The wave function is localized about one of the two coordinates
transverse to the magnetic field (the $x$-coordinate),
\be
{\bf x}_\perp = a\hat x.
\ee
The corresponding transverse momentum is
\ba
{\bf P}_\perp &=& e{\bf x}_\perp \times {\bf B}\nn
      &=& -a\,eB\hat y \;\equiv\; -{a\over\lambda_B^2}\hat y.
\ea
In a frame where $P = 0$, the positive-energy wave functions are
\be
u^{(-1)}_{n,a}({\bf x}_\perp) = {1\over L (2E_{0n})^{1/2}}
\left[\begin{array}{c}
0 \\ 
(E_{0n}+m_e)^{1/2}\,\phi_n(x-a) \\
-i(E_{0n}-m_e)^{1/2}\,\phi_{n-1}(x-a) \\
0 \\
\end{array}
\right]\qquad(n \geq 0)
\ee
and
\be
u^{(+1)}_{n,a}({\bf x}_\perp) = {1\over L (2E_{0n})^{1/2}}
\left[\begin{array}{c}
(E_{0n}+m_e)^{1/2}\,\phi_{n-1}(x-a) \\
0 \\ 
0 \\ 
i(E_{0n}-m_e)^{1/2}\,\phi_n(x-a) \\
\end{array}
\right]\qquad(n \geq 1),
\ee
where $L^3$ is the normalization volume.  The $x$-dependence of
the wave function is
\be
\phi_n(x-a) = {1\over (2^n n!)^{1/2}}
\left({eB\over \pi}\right)^{1/4}\,
\exp\left[-{1\over 2}\left({x-a\over\lambda_B}\right)^2\right]
\,H_n\left({x-a\over\lambda_B}\right),
\ee
where $H_n$ is the Hermite polynomial.  
In a frame boosted to a momentum $P$ parallel to ${\bf B}$, the wave
functions become
\be
u^{(-1)}_{n,a}({\bf x}_\perp) = {1\over 2L [EE_{0n}(E+E_{0n})]^{1/2}}
\left[\begin{array}{c}
-iP(E_{0n}-m_e)^{1/2}\,\phi_{n-1}\\
(E+E_{0n})(E_{0n}+m_e)^{1/2}\,\phi_n \\
-i(E+E_{0n})(E_{0n}-m_e)^{1/2}\,\phi_{n-1} \\
-P(E_{0n}+m_e)^{1/2}\,\phi_n\\
\end{array}
\right],
\ee
\be
u^{(+1)}_{n,a}({\bf x}_\perp) = {1\over 2L [EE_{0n}(E+E_{0n})]^{1/2}}
\left[\begin{array}{c}
(E+E_{0n})(E_{0n}+m_e)^{1/2}\,\phi_{n-1} \\
-iP(E_{0n}-m_e)^{1/2}\,\phi_n\\
P(E_{0n}+m_e)^{1/2}\,\phi_{n-1}\\
i(E+E_{0n})(E_{0n}-m_e)^{1/2}\,\phi_n \\
\end{array}
\right].
\ee
The positron wave functions are given by
\be
v^{(-1)}_{n,a}({\bf x}_\perp) = {1\over 2L [EE_{0n}(E+E_{0n})]^{1/2}}
\left[\begin{array}{c}
i(E+E_{0n})(E_{0n}-m_e)^{1/2}\,\phi_{n-1} \\
-P(E_{0n}+m_e)^{1/2}\,\phi_n\\
iP(E_{0n}-m_e)^{1/2}\,\phi_{n-1}\\
(E+E_{0n})(E_{0n}+m_e)^{1/2}\,\phi_n \\
\end{array}
\right],
\ee
\be
v^{(+1)}_{n,a}({\bf x}_\perp) = {1\over 2L [EE_{0n}(E+E_{0n})]^{1/2}}
\left[\begin{array}{c}
P(E_{0n}+m_e)^{1/2}\,\phi_{n-1}\\
-i(E+E_{0n})(E_{0n}-m_e)^{1/2}\,\phi_n \\
(E+E_{0n})(E_{0n}+m_e)^{1/2}\,\phi_{n-1} \\
iP(E_{0n}-m_e)^{1/2}\,\phi_n\\
\end{array}
\right].
\ee
In these expressions, $E_{0n}$ is the 
energy at $P=0$, and $E = E(P)$ (eq. [\ref{eq:disrele}]).

The wavefunctions of the two photons (four-momenta $k_i = 
(\omega_i,\, {\bf k}_i)$, polarization vectors 
$\varepsilon^\mu_i$, $i = 1,2$) are normalized in the usual way,
\be
A^\mu_i(x) \;=\;
 {e^{ik_i\cdot x}\over (2\omega_i)^{1/2} L^{3/2}}
\;\varepsilon^\mu_i.
\ee

\subsection{Case 1: Head on Collision, Center-of-Momentum Frame}
\label{eq:headon}

We consider the photon collision not too far
above the threshold for pair creation, in a range of energies
where the outgoing electron and positron must be confined to their
lowest Landau states, $n_- = n_+ = 0$.  We work in the center-of-momentum frame
where
\be
\omega_1 = k_{1,z} = \omega;
\qquad
\omega_2 = -k_{1,z} = \omega.
\ee
The energies and momenta of the outgoing electron and positron are then 
\be
E_- = E_+ = \omega\qquad P_- = -P_+ = \sqrt{\omega^2-m_e^2}.
\ee

We calculate the matrix elements
\be
S_{fi} = S_{fi}[1] + S_{fi}[2];
\ee
\ba
S_{fi}[1] &=& ie^2 \int d^4x \int d^4 x' 
\left[\bar\psi_-^{(-1)}(x')\right]_{n_-=0}\gamma_\mu A^\mu_2(x') 
G_F(x'-x)\gamma_\nu A_1^\nu(x) \left[\psi_+^{(-1)}(x)\right]_{n_+=0};\nn
S_{fi}[2] &=& ie^2 \int d^4x \int d^4 x' 
\left[\bar\psi_-^{(-1)}(x')\right]_{n_-=0} \gamma_\mu A^\mu_1(x') 
G_F(x'-x)\gamma_\nu A_2^\nu(x)\left[\psi_+^{(-1)}(x)\right]_{n_+=0},\nn
\ea
using the real-space representation of the electron propagator
\ba\label{eq:gf}
G_F(x'-x) &=& L^2\int{da_I\over 2\pi \lambda_B^2}\int {dP_I\over 2\pi}
\sum_{n_I=0}^\infty
\biggl[ 
-i\theta(t'-t)\sum_{\sigma_I} u_{n_I}^{(\sigma_I)}({\bf x}') 
\bar u_{n_I}^{(\sigma_I)} ({\bf x}) e^{-iE_I(t'-t)}
e^{i{\bf p}\cdot({\bf x}'- {\bf x})} \nn
&+& i\theta(t-t')\sum_{\sigma_I} 
v_{n_I}^{(\sigma_I)}({\bf x}')\bar v_{n_I}^{(\sigma_I)}
({\bf x}) e^{iE_I(t'-t)}e^{-i{\bf p}\cdot({\bf x}'- {\bf x})}
\biggr].
\ea
Here $\bar\psi = \psi^\dagger\gamma_0 =
(\psi^*)^T\gamma_0$, where $*$ denotes complex conjugation and
$T$ the transpose.  
Performing the integrals over $t$, $t'$ yields
a factor $2\pi\delta(E_+ + E_- - \omega_1-\omega_2)
(E_- - \omega_2\mp E_I)^{-1}$ for each of the two terms in $G_F$,
and we have
\ba\label{eq:matrix}
S_{fi}[1] &=& -{i e^2\over 2\omega}{1\over L^3}
\left({L\over 2\pi}\right)^2\,2\pi\delta(E_+ + E_--2\omega)\nn
&&\qquad\times\int dP_I \int {da_I\over\lambda_B^2}\;
\sum_{\sigma_I}\int\left({I_1^{(\sigma_I)}I_2^{(\sigma_I)}\over
  E_- - \omega - E_I} + {I_3^{(\sigma_I)}I_4^{(\sigma_I)}\over
  E_- - \omega + E_I}\right),
\ea
where
\ba
I_1^{(\sigma_I)} &=& \int d^3x'
\left[\bar u^{(-1)}_{0,a_-}({\bf x}'_\perp)\gamma_\mu
 u_{1,a_I}^{(\sigma_I)}({\bf x}_\perp')\right]\varepsilon^\mu_2
e^{i({\bf k}_2-{\bf p}_- +{\bf p}_I)\cdot{\bf x}'};\nn
I_2^{(\sigma_I)} &=& \int d^3x
\left[\bar u^{(\sigma_I)}_{1,a_I}({\bf x}_\perp)\gamma_\nu
 v_{0,a_+}^{(-1)}({\bf x}_\perp)\right]\varepsilon^\nu_1
e^{i({\bf k}_1-{\bf p}_+-{\bf p}_I)\cdot{\bf x}};\nn
I_3^{(\sigma_I)} &=& \int d^3x'
\left[\bar u^{(-1)}_{0,a_-}({\bf x}'_\perp)\gamma_\mu
 v_{1,a_I}^{(\sigma_I)}({\bf x}_\perp')\right]\varepsilon^\mu_2
e^{i({\bf k}_2-{\bf p}_+-{\bf p}_I)\cdot{\bf x}'};\nn
I_4^{(\sigma_I)} &=& \int d^3x
\left[\bar v^{(\sigma_I)}_{1,a_I}({\bf x}_\perp)\gamma_\nu
 v_{0,a_+}^{(-1)}({\bf x}_\perp)\right]\varepsilon^\nu_1
e^{i({\bf k}_1-{\bf p}_++{\bf p}_I)\cdot{\bf x}}.
\ea
These integrals involve a matrix 
\be
\gamma_0\gamma_\nu\,\varepsilon_i^\mu  = 
\left(\begin{array}{cc}
1 & 0 \\
0 & -1 \\
\end{array}\right)\,\varepsilon_i^t +
\left(\begin{array}{cc}
0 & \sigma_x \\
\sigma_x & 0 \\
\end{array}\right)\,\varepsilon_i^x +
\left(\begin{array}{cc}
0 & \sigma_y \\
\sigma_y & 0 \\
\end{array}\right)\,\varepsilon_i^y +
\left(\begin{array}{cc}
0 & \sigma_z \\
\sigma_z & 0 \\
\end{array}\right)\,\varepsilon_i^z,
\ee
where $\sigma_i$ are the $2\times 2$ Pauli matrices.  Expanding, one finds
\be
\gamma_0\gamma_\nu\,\varepsilon_i^\mu  = 
\left(\begin{array}{cccc}
0 & 0 & \varepsilon_i^z & \varepsilon_i^- \\
0 & 0 & \varepsilon_i^+ & -\varepsilon_i^z\\
\varepsilon_i^z & \varepsilon_i^- & 0 & 0\\
\varepsilon_i^+ & -\varepsilon_i^z & 0  & 0\\
\end{array}\right),
\ee
where $\varepsilon_i^\pm = \varepsilon_i^x \pm i\varepsilon_i^y$.

It will be noted that only the Landau level $n_I=1$ in the propagator
contributes to the matrix element.  That is because
\be
\bar u^{(-1)}_{0,a_1}\gamma_\mu  u_{0,a_2}^{(-1)}\varepsilon^\mu_i
\;\propto\; \varepsilon_i^z
\ee
vanishes when the two photons are incident parallel to ${\bf B}$,
and because integrals of the form
\be
\int d^2x_\perp
\phi_0(x-a_1)\phi_m(x-a_2) e^{i({\bf k}_i-
{\bf p}_- +{\bf p}_I)_\perp\cdot{\bf x}_{\perp}}
\;\propto\; \left|{\bf k}_{i,\perp}\right|^m
\ee
also vanish when $m\geq 1$.  We are therefore left with evaluating
integrals of the form
\be
\int d^3x
\phi_0(x-a_1)\phi_0(x-a_2) e^{i({\bf k}_i-
{\bf p}_1 +{\bf p}_2)\cdot{\bf x}}
= \left({2\pi\over L}\right)^2\,\delta\left({a_1-a_2\over\lambda_B^2}\right)
\,\delta(k_{i,z} - P_1 + P_2),
\ee
where the result assumes ${\bf k}_{i,\perp} = 0$.
For example, the integrals $I_1$, $I_2$ take the form
\be
I_1^{(\sigma_I)} = {F_1^{(\sigma_I)}\varepsilon^+_2\over
4[E_-E_Im_eE_{0n}(E_-+m_e)(E_I+E_{0n})]^{1/2}}
\,\left({2\pi\over L}\right)^2\delta\left({a_--a_I\over\lambda_B^2}\right)
\delta(k_{2,z}- P_- + P_I),
\ee
\be
I_2^{(\sigma_I)} = {F_2^{(\sigma_I)}\varepsilon^-_1\over
4[E_+E_Im_eE_{0n}(E_++m_e)(E_I+E_{0n})]^{1/2}}
\,\left({2\pi\over L}\right)^2\delta\left({a_I-a_+\over\lambda_B^2}\right)
\delta(k_{1,z}- P_+ - P_I),
\ee
where 
\be
E_{0n} \equiv (m_e^2 + 2eB)^{1/2};\qquad E_I \equiv
\left[(\omega+P_-)^2 + m_e^2 + 2eB\right]^{1/2}
\ee
and
\ba
F_1^{(-1)} &=&
 i\Bigl[P_-P_I-(E_-+m_e)(E_I+E_{0n})\Bigr](2m_e)^{1/2}(E_{0n}-m_e)^{1/2};\nn
F_1^{(+1)} &=&
 \Bigl[(E_-+m_e)P_I - (E_I+ E_{0n})P_-\Bigr](2m_e)^{1/2}(E_{0n}+m_e)^{1/2};\nn
F_2^{(-1)} &=&
 i\Bigl[(E_++m_e)P_I - (E_I+ E_{0n})P_+\Bigr](2m_e)^{1/2}(E_{0n}-m_e)^{1/2};\nn
F_2^{(+1)} &=&
 \Bigl[P_+P_I-(E_++m_e)(E_I+E_{0n})\Bigr](2m_e)^{1/2}(E_{0n}+m_e)^{1/2}.
\ea
Summing gives
\ba\label{eq:isum1}
I_1^{(-1)}I_2^{(-1)} + I_1^{(+1)}I_2^{(+1)}
&=& \varepsilon_1^-\varepsilon_2^+\,{m_e(P_I-P_\pm)\over 2E_\pm E_I}\,
\left({2\pi\over L}\right)^4\,\delta\left({a_--a_I\over\lambda_B^2}\right)
\delta\left({a_+-a_I\over\lambda_B^2}\right)\nn
&&\qquad\,\times\delta(\omega+P_--P_I)\,\delta(\omega - P_+-P_I).
\ea
and
\ba\label{eq:isum2}
I_3^{(-1)}I_4^{(-1)} + I_3^{(+1)}I_4^{(+1)}
&=& \varepsilon_1^-\varepsilon_2^+\,{m_e(P_I+P_\pm)\over 2E_\pm E_I}\,
\left({2\pi\over L}\right)^4\,\delta\left({a_-+a_I\over\lambda_B^2}\right)
\delta\left({a_++a_I\over\lambda_B^2}\right)\nn
&&\qquad\,\times\delta(\omega+P_-+P_I)\,\delta(\omega - P_++P_I).
\ea
We have used the notation $P_\pm = P_- = -P_+$ and $E_\pm = E_+ = E_-$,
and substituted $k_{1,z} = -k_{2,z} = \omega$.

The first term in the matrix element is obtained by substituting
eqs. (\ref{eq:isum1}) and (\ref{eq:isum2}) into eq. (\ref{eq:matrix}),
\be
S_{fi}[1] \;=\; -{ie^2\over 2L^5}\,{m_e \varepsilon_1^-\varepsilon_2^+
\over \omega [(\omega + P_-)^2 + m_e^2 + 2eB]}\;\;
(2\pi)^3\,\delta(E_-+E_+-2\omega)\,\delta(P_+ + P_-)\,
\delta\left({a_--a_+\over\lambda_B^2}\right).
\ee
The second contribution to the matrix element is obtained by 
exchanging the two photons, which corresponds to taking
$P_I \rightarrow P_- - \omega$ in the denominator,
\be
S_{fi}[2] \;=\; {ie^2\over 2L^5}\,{m_e \varepsilon_1^+\varepsilon_1^-
\over \omega [(\omega - P_-)^2 + m_e^2 + 2eB]}\;\;
(2\pi)^3\,\delta(E_-+E_+-2\omega)\,\delta(P_+ + P_-)\,
\delta\left({a_--a_+\over\lambda_B^2}\right).
\ee

The total cross section for $\gamma + \gamma \rightarrow e^+ + e^-$,
with the electron and positron both in the lowest Landau state, is
obtained from
\be
(1-\cos\theta_{12})\,\sigma = 
{L^3\over T}
\int L{da_-\over 2\pi\lambda_B^2}\,\int L{da_+\over 2\pi\lambda_B^2}\,
\int L{dp_-\over 2\pi}\,\int L{dp_+\over 2\pi}\,
\Bigl|S_{fi}[1]+S_{fi}[2]\Bigr|^2.
\ee
Here $\cos\theta_{12}= -1$ is the cosine of the collision angle
between the two photons, and $T$ is a normalization time.  The square
of a delta function is handled with the substitution
$\left[(2\pi)\,\delta(E_-+E_+-2\omega)\right]^2 \rightarrow
T\,(2\pi)\,\delta(E_-+E_+-2\omega)$,
$\left[(2\pi)\,\delta(P_+ + P_-)\right]^2\rightarrow L\,
(2\pi)\,\delta(P_+ + P_-)$, and so on, yielding
\be
\sigma = {e^4 \over 32\pi \omega^2}\,\left({B\over\BQ}\right)
\left|{dE_\pm\over dP_\pm}\right|^{-1}\,
\left|{\varepsilon_2^+\varepsilon_1^- m_e^2\over
(\omega + P_\pm)^2 + m_e^2 + 2eB} -
{\varepsilon_1^+\varepsilon_2^- m_e^2\over
(\omega - P_\pm)^2 + m_e^2 + 2eB}\right|^2.
\ee
Here $dE_\pm/dP_\pm = V_\pm$
is the speed of the outgoing electron and positron.
The coefficient in front is readily expressed in terms of the Thomson
cross section
\be
\sigma_T = {8\pi\over 3}\left({e^2\over 4\pi m_e}\right)^2.
\ee
Near threshold, the cross section vanishes if the two photons
have the same linear polarization,
\be
{\sigma\over\sigma_T} = {3B/\BQ\over 16(1+B/\BQ)^2}\,
\left({V_\pm\over c}\right)^{-1}\,
\left(\varepsilon^x_1\varepsilon^y_2-\varepsilon^x_2\varepsilon^y_1\right)^2
\qquad(\omega\rightarrow m_e).
\ee

\subsection{Case 2: Photon 1 Near Threshold for Pair 
Creation}\label{s:case2}

We now turn to the case where one photon (labeled 1) is just
below threshold for pair creation.  We choose a frame in
which $\omega_1 = 2m_e -\Delta E$,
$k_{1,z} = 0$, and $\omega_2 = k_{2,z}$.
We focus on the case where there is a near resonance in the matrix
element, in the sense that the virtual electron is nearly on mass shell.
Simple kinematic considerations show that this corresponds to one
outgoing particle in the first Landau state, with the virtual
particle in level $n_I = 0$ (\S \ref{s:gxcol}).  The final-state particle
in state $n=1$ is the one that shares a vertex with photon 2 in 
Fig. \ref{f:diagram}.  The final states with $n_- = 1$, $n_+ = 0$ and
$n_- = 0$, $n_+ = 1$ are distinguishable, and so we may calculate
the cross section for the electron in the excited state, and then
multiply by a factor 2 to obtain the total cross section.  The excited particle
also comes in two spin states $\sigma = \pm 1$, which must be summed over.

The matrix elements to be calculated are therefore
\be\label{eq:matrixb}
S_{fi}^{(\sigma_-)} = ie^2 \int d^4x \int d^4 x' 
\left[\bar\psi_-^{(\sigma_-)}(x')\right]_{n_-=1} \gamma_\mu A^\mu_2(x') 
G_F(x'-x)\gamma_\nu A_1^\nu(x) \left[\psi_+^{(-1)}(x)\right]_{n_+=0}
\qquad (\sigma_- = \pm 1),
\ee
and the cross section is given by
\be
(1-\cos\theta_{12})\sigma = 2{L^3\over T}
\int L{da_-\over 2\pi\lambda_B^2}\,\int L{da_+\over 2\pi\lambda_B^2}\,
\int L{dp_-\over 2\pi}\,\int L{dp_+\over 2\pi}\,
\left(\Bigl|S_{fi}^{(-1)}\Bigr|^2 + \Bigl|S_{fi}^{(+1)}\Bigr|^2\right),
\ee
where in this case $\cos\theta_{12} = 0$.  The condition for a near
resonance is that $E_I \simeq E_- - \omega_2$, and so we need
only include the term in expression (\ref{eq:gf}) for $G_F(x'-x)$ 
involving the positive-energy wavefunctions.  The matrix elements
(\ref{eq:matrixb}) become 
\be\label{eq:scat1}
S_{fi}^{(\sigma_-)} = -{ie^2\over 2 (\omega_1\omega_2)^{1/2}}
{1\over L^3}\left({L\over 2\pi}\right)^2 2\pi \delta(E_+ + E_- - 
\omega_1 - \omega_2)\int {da_I\over \lambda_B^2}\int dp_I
{I_1^{(\sigma_-)}I_2\over E_--\omega_2 - E_I},
\ee
where
\ba\label{eq:scat2}
I_1^{(\sigma_-)} &=& \int d^3x'
\left[\bar u^{(\sigma_-)}_{1,a-}({\bf x}'_\perp)\gamma_\mu
 u_{0,a_I}^{(-1)}({\bf x}_\perp')\right]\varepsilon^\mu_2
e^{i({\bf k}_2-{\bf p}_- +{\bf p}_I)\cdot{\bf x}'};\nn
I_2 &=& \int d^3x
\left[\bar u^{(-1)}_{0,a_I}({\bf x}_\perp)\gamma_\nu
 v_{0,a_+}^{(-1)}({\bf x}_\perp)\right]\varepsilon^\nu_1
e^{i({\bf k}_1-{\bf p}_+-{\bf p}_I)\cdot{\bf x}}.
\ea
Near threshold for pair creation, one has $E_+ = E_I \simeq m_e$,
$E_- \simeq m_e + eB/m_e$,  and $\omega_2\simeq eB/m_e$.
The above integrals may then be evaluated as
\be\label{eq:scat3}
I_1^{(\sigma_-)} = {F_1^{(\sigma_-)}\varepsilon^-_2\over
2[E_-E_{0n}(E_{0n}+m_e)(E_-+E_{0n})]^{1/2}}
\,\left({2\pi\over L}\right)^2\delta\left({a_--a_I\over\lambda_B^2}\right)
\delta(k_{2,z}- P_- + P_I);
\ee
\be\label{eq:scat4}
I_2 = \varepsilon_1^z\,
\exp\left(-{\lambda_B^2 \omega_1^2\over 4}\right)\,
\;\left({2\pi\over L}\right)^2
\delta\left(\omega_1-{a_+-a_I\over\lambda_B^2}\right)\delta(P_-+P_+),
\ee
where
\be\label{eq:scat5}
F_1^{(-1)} = i(E_- + E_{0n})(2eB)^{1/2}; \qquad
F_1^{(+1)} = P_-(E_{0n}+m_e),
\ee
and  we have chosen the momentum of photon 1 to point in the $y$-direction.
The cross section is proportional to 
\be
{|F_1^{(-1)}|^2 + |F_1^{(+1)}|^2 \over 4E_-E_{0n}(E_{0n}+m_e)(E_-+E_{0n})}
\simeq {E_--m_e\over 2E_-} = {B/\BQ\over 2(1+B/\BQ)}.
\ee

Substituting eqs. (\ref{eq:scat2})-(\ref{eq:scat4}) into (\ref{eq:scat1}),
and averaging over the spins of photon 2, one finds 
\be
\sigma \;=\; {e^4\over 16\pi (\omega_1-2E_+)^2}\,
\exp\left(-{\omega_1^2\over 2eB}\right)
\left({B/\BQ\over 1+B/\BQ}\right)\,
\left|{dP_+\over d(E_-+E_+)}\right|_{P_-+P_+=\omega_2}\,
(\varepsilon_1^z)^2.
\ee
As expected, the cross section is non-vanishing only if photon 1
is polarized parallel to ${\bf B}$.
The last factor is determined by differentiating
\be
E_- + E_+ = (P_-^2 + m_e^2 + 2eB)^{1/2} + (P_+^2 + m_e^2)^{1/2}
\ee
with 
$P_- = \omega_2 - P_+$.  Near threshold,
\be
\left|{dP_+\over d(E_-+E_+)}\right|_{P_-+P_+=\omega_2} 
\simeq {1 + B/\BQ\over B/\BQ},
\ee
and we have
\be
\sigma \;=\; {e^4\over 16\pi (\omega_1-2E_+)^2}\,
\exp\left(-{\omega_1^2\over 2eB}\right)(\varepsilon_1^z)^2.
\ee
The cross section can then be expressed in terms of $\sigma_T$
(eq. [\ref{eq:sigresxg}]).

\end{appendix}


\end{document}